# Linear Shift-Register Synthesis for Multiple Sequences of Varying Length

Georg Schmidt, *Student Member, IEEE*, and Vladimir R. Sidorenko, *Member, IEEE*
Department of Telecommunications and Applied Information Theory, University of Ulm, Germany
{georg.schmidt,vladimir.sidorenko}@uni-ulm.de

*Abstract*— The problem of finding the shortest linear shift-register capable of generating $t$ finite length sequences over some field $\mathbb{F}$ is considered. A similar problem was already addressed by Feng and Tzeng. They presented an iterative algorithm for solving this multi-sequence shift-register synthesis problem, which can be considered as generalization of the well known Berlekamp–Massey algorithm. The Feng–Tzeng algorithm works indeed, if all $t$ sequences have the same length. This paper focuses on multi-sequence shift-register synthesis for generating sequences of varying length. It is exposed, that the Feng–Tzeng algorithm does not always give the correct solution in this case. A modified algorithm is proposed and formally proved, which overcomes this problem.

## I. INTRODUCTION

In [1], [2], and [3], algorithms for synthesizing a minimum length linear shift-register for generating multiple sequences of finite length over some field are proposed. These algorithms can be seen as generalizations of a well known algorithm from Berlekamp [4] and Massey [5], which forms the backbone for efficiently solving a variety of technical problems.

Linear shift-register synthesis for multiple sequences can be applied in the field of coding theory and provides methods for efficiently decoding of cyclic codes, interleaved codes, concatenated codes and some other code constructions for channels with independent errors as well as for channels with memory. Problems for which shift-register synthesis for multiple sequences can be beneficial are described e.g., in [2], [6], [7], [8], [9], and [10]. Linear shift-register synthesis for multiple sequences of *varying length* can efficiently be applied for example in interleaved schemes with codes of different rates, error and erasure correction, in some channels with memory and some other applications (see, e.g. [6], [9], [11], [12], and [13]). There exist some generalizations and variations of the problem. Sequences over rings are considered in [14], and [15], periodic sequences of infinite length are analyzed in [16].

In this paper, we focus on the problem of finding the shortest possible shift-register capable of generating multiple sequences of varying length over some field $\mathbb{F}$. The Feng–Tzeng algorithm proposed in [1], and [2] solves the shift-register synthesis problem for multiple sequences *of equal length*. Moreover, it is stated in [2], that the Feng–Tzeng algorithm is also suitable for solving the multi-sequence shift-register synthesis problem for *varying length* sequences. However, it turns out, that this is not always true. We illustrate by means of a simple counter-example, that the *Fundamental Iterative Algorithm* (FIA) described in [2] fails to give the shortest possible shift-register in some cases, when applied to multiple sequences of varying length. Consequently, also the Feng–Tzeng algorithm derived from the FIA is not applicable for sequences of varying length. We explain, why this problem occurs and how the FIA needs to be modified in order to work correctly for arbitrary varying length sequences. Based on this modification, we derive an efficient shift-register synthesis algorithm, whose structure is similar to the Berlekamp–Massey algorithm and the Feng–Tzeng algorithm respectively.

This paper is intended to be suitable not only for experts in the field of shift-register synthesis, but also for a broad spectrum of readers. To give readers which are not familiar with [1] and [2] the opportunity to understand conveniently the basic ideas behind efficient multi-sequence shift-register synthesis, this paper is structured such that we completely derive and proof all necessary components of the algorithm step by step, closely following the ideas of Feng and Tzeng. To aid experienced readers already familiar with [2], we use the notation introduced in [2] whenever it is expedient. We explicitly point out at which parts of the algorithm we meet problems with varying length sequences, and explain, what modifications are necessary to guarantee that the shortest possible shift-register can always be obtained.

## II. VARIANTS OF SHIFT-REGISTER SYNTHESIS

Shift-register synthesis problems occur in various applications in several variants. Notwithstanding the fact, that we are effectively interested in the problem of shift-register synthesis for multiple sequences of varying length, we start with briefly discussing several variants of shift-register synthesis problems to explain an illustrate several aspects which will be important in the following.

### A. Single-Sequence Shift-Register Synthesis

Consider a sequence $\mathcal{S} = \{S_i\}_{i=1}^{N} = S_1, \ldots, S_N$ of $N$ symbols over some field $\mathbb{F}$. Now, the task is to find the shortest

The work of Vladimir R. Sidorenko and Georg Schmidt is supported by Deutsche Forschungsgemeinschaft (DFG), Germany, under project BO 867/14, and project BO 867/15 respectively. Some of the results presented in this paper will also be presented at the IEEE International Symposium on Information Theory, Seattle, USA, 2006. This work has been submitted to the IEEE for possible publication. Copyright may be transferred without notice, after which this version will be superseded.

possible linear feedback shift-register like depicted in Fig. 1, such that it generates the complete sequence if it is initialized by the subsequence $\{S_i\}_{i=1}^{l}$. In other words, the problem of synthesizing a minimum length linear feedback shift-register is to find the smallest possible positive integer $l$, and a connection polynomial $\sigma(x) = 1 + \sigma_1 x + \cdots + \sigma_l x^l$ such that

$$S_j = -\sum_{i=1}^{l} \sigma_i S_{j-i},\ \forall j = l+1, \ldots, N\ .$$

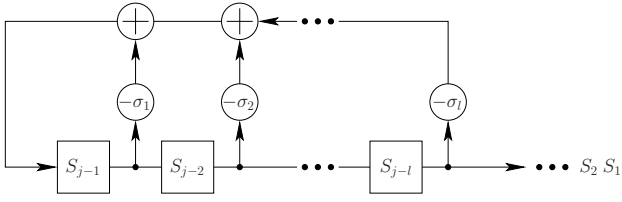

Fig. 1. Shift-register synthesis for a single sequence

Note, that some of the coefficients $\sigma_i$ may be zero, and therefore the degree of $\sigma(x)$ can be smaller than the length of the shift-register. In this case, some of the rightmost memory elements of the shift-register have no taps, and therefore the first elements of $S_1, \ldots, S_N$ have no influence on the feedback path of the shift-register. Consequently, the length $l$ of a shift-register can be very large, even if the degree of $\sigma(x)$ is small.

**Example 1** *Consider the sequence*

$$\{S_i\}_{i=1}^{10} = 0,0,0,0,1,1,1,1,1,1$$

*with elements from the binary field* $\mathbb{F}_2$. *This sequence can be generated by the minimum length shift-register depicted in Fig. 2. This shift-register has length* $l = 5$, *even though the connection polynomial* $\sigma(x) = 1 + x$ *has degree* $\deg(\sigma(x)) = 1$.

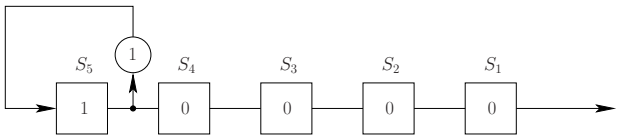

Fig. 2. Minimum length shift-register for a binary sequence

Formally we can state the classical single-sequence shift-register synthesis problem as follows:

**Problem 0** *Let* $\mathcal{S} = \{S_i\}_{i=1}^{N}$ *be a sequence of length* $N$. *Then, the* single-sequence shift-register synthesis problem *is formulated as follows: Find the smallest integer* $l$, *and a polynomial* $\sigma(x) = 1 + \sigma_1 x + \cdots + \sigma_l x^l$, $\deg(\sigma(x)) \leq l$, *such that*

$$S_j = -\sum_{i=1}^{l} \sigma_i S_{j-i},\ \forall j = l+1, \ldots, N\ .$$

At this, the integer $l$ is called the *linear complexity* of the sequence $\mathcal{S}$.

Problem 0 is the classical shift-register problem considered by Berlekamp [4] and Massey [5]. It serves as starting point for considering several multi-sequence shift-register problems in the following.

### B. Multi-Sequence Shift-Register Synthesis

Now, we generalize Problem 0 by considering several sequences simultaneously. More precisely, we consider $t$ different sequences $\mathcal{S}^{(h)} = \{S_i^{(h)}\}_{i=1}^{N}$, $h = 1, \ldots, t$ of length $N$, with elements from some field $\mathbb{F}$. Now again, we like to find the shortest possible linear feedback shift-register, which is able to generate all sequences $\mathcal{S}^{(h)}$, $h = 1, \ldots, t$, from the corresponding initial subsequences $\{S_i^{(h)}\}_{i=1}^{l}$. Formally we state the multi-sequence shift-register synthesis problem in the following way:

**Problem 1** *Let* $\mathcal{S} = \{S_i^{(h)}\}_{i=1}^{N}$, $h = 1, \ldots, t$, *be* $t$ *sequences of length* $N$. *Then, the* multi-sequence shift-register synthesis problem *is formulated as follows: Find the smallest integer* $l$, *and a polynomial* $\sigma(x) = 1 + \sigma_1 x + \cdots + \sigma_l x^l$, $\deg(\sigma(x)) \leq l$, *such that*

$$S_j^{(h)} = -\sum_{i=1}^{l} \sigma_i S_{j-i}^{(h)},\ \forall j = l+1, \ldots, N,\ h = 1, \ldots, t\ .$$

An algorithm for solving Problem 1 is developed by Feng and Tzeng in [1], and [2]. A similar algorithm is derived in [3] from a lattice base reduction algorithm. Another algorithm for the multi-sequence shift-register synthesis problem is conjectured by Massey and proved in [17].

For $t = 1$, Problem 1 reduces to Problem 0. In this sense, the classical single-sequence shift-register synthesis problem is a special case of the multi-sequence shift-register problem described by Problem 1.

### C. Varying Length Multi-Sequence Shift-Register Synthesis

The multi-sequence shift-register synthesis problem can further be generalized by allowing the $t$ sequences $\mathcal{S}^{(h)}$ to have different length $N_h \leq N$, where $N = \max\{N_h\}$, and $h = 1, \ldots, t$.

**Problem 2** *Let* $\mathcal{S}^{(h)} = \{S_i^{(h)}\}_{i=1}^{N_h}$, $h = 1, \ldots, t$, *be* $t$ *sequences of length* $N_h \leq N$, *where* $N = \max\{N_h\}$. *Then, the* varying length multi-sequence shift-register problem *is formulated as follows: Find the smallest integer* $l$, *and a polynomial* $\sigma(x) = 1 + \sigma_1 x + \cdots + \sigma_l x^l$, $\deg(\sigma(x)) \leq l$ *such that*

$$S_j^{(h)} = -\sum_{i=1}^{l} \sigma_i S_{j-i}^{(h)},\ \forall j = l+1, \ldots, N_h,\ h = 1, \ldots, t\ .$$

It is stated in [2], that the Feng–Tzeng algorithm can also be applied for sequences of varying length, i.e., for solving Problem 2. However, it turns out that this algorithm does not generally obtain a correct solution for Problem 2. We demonstrate this in Section IV, and explain, how the algorithm can be modified and extended such that it is guaranteed to find a correct solution for Problem 2.

## III. Sequences with Wild Cards

To treat Problem 1 and Problem 2 in a unified way, we reduce Problem 2 to Problem 1 by virtually prolonging the sequences which are shorter than $N = \max\{N_h\}$ by appending $N - N_h$ special symbols which we call *wild cards*. Formally we describe this prolongation by the following definition:

**Definition 1**
Let $N > N_h$, and denote by

$$\mathcal{S}|\mathcal{X} = \{S_1, S_2, \ldots, S_{N_h}, X_1, X_2, \ldots X_{N-N_h}\}$$

the concatenation of the sequence $\mathcal{S} = \{S_i\}_{i=1}^{N_h}$ with elements $S_i \in \mathbb{F}$ and the sequence $\mathcal{X} = \{X_i\}_{i=1}^{N-N_h}$ whose elements are special characters called wild cards. These wild cards are placeholders for field elements, i.e., they can be replaced by arbitrary elements from $\mathbb{F}$. In this way, $\mathcal{S}|\mathcal{X}$ describes a set of $|\mathbb{F}|^{N-N_h}$ different sequences with elements from $\mathbb{F}$. Each particular choice of the wild cards selects one particular sequence from this set.

In shift-register synthesis problems, the meaning of the wild cards is, that each of these placeholders can be replaced with a field element chosen by the synthesis algorithm to optimize, in some sense, the resulting solution of the synthesis problem. Usually, the wild cards are chosen such, that the linear complexity of the fully specified sequence is minimized. In other words, the wild cards should be specified such, that the length of the shift-register capable of generating the sequence is as small as possible. In this sense, the wild cards are a means of prolonging a sequence without increasing its linear complexity. To illustrate this, we consider the following example:

**Example 2** *Assume, that the wild cards of the binary sequence*

$$\mathcal{S} = 1, 0, 1, 0, 1, 0, X_1, X_2$$

*shall be chosen such, that the linear complexity of the fully specified sequence is minimized. Here the unique solution is $X_1 = 1$ and $X_2 = 0$.*

Using the concept of wild cards, we are now able to restate Problem 2 in such a way, that we formally consider all $t$ sequences having the same length $N$:

**Problem 2'** *Let $\mathcal{S}^{(h)} = \{\tilde{S}_i^{(h)}\}_{i=1}^{N} = \{S_i^{(h)}\}_{i=1}^{N_h} | \{X_i\}_{i=1}^{N-N_h}$, $h = 1, \ldots, t$, be $t$ sequences of length $N = \max\{N_h\}$, where the last $N - N_h$ symbols of every sequence are wild cards. Then, the* varying length multi-sequence shift-register problem *is formulated as follows: Among all possible choices of the wild cards, find the smallest integer $l$, and a polynomial $\sigma(x) = 1 + \sigma_1 x + \cdots + \sigma_l x^l$, $\deg(\sigma(x)) \leq l$, such that*

$$\tilde{S}_j^{(h)} = -\sum_{i=1}^{l} \sigma_i \tilde{S}_{j-i}^{(h)}, \;\forall j = l+1, \ldots, N, \; h = 1, \ldots, t \;.$$

**Remark:** Since the linear complexity of a sequence cannot be decreased by prolonging the sequence, any solution of Problem 2 is also a solution of Problem 2'. Vice versa, any solution of Problem 2' also solves Problem 2. Hence, the problems 2 and 2' are equivalent.

## IV. Minimum Length Shift-register synthesis by a Fundamental Iterative Algorithm

The concept of wild cards can be applied to reformulate the multi-sequence shift-register problem as a problem from linear algebra. For this purpose, we create the matrix

$$\boldsymbol{S} = \begin{pmatrix} S_1^{(1)} & S_2^{(1)} & \ldots & S_{N_1-1}^{(1)} & S_{N_1}^{(1)} & X_1^{(1)} & \ldots & \ldots & X_{N-N_1}^{(1)} \\ \vdots & \vdots & & & & & & & \vdots \\ S_1^{(\ell)} & S_2^{(\ell)} & \ldots & \ldots & \ldots & \ldots & \ldots & \ldots & S_N^{(\ell)} \\ \vdots & \vdots & & & & & & & \vdots \\ S_1^{(t)} & S_2^{(t)} & \ldots & \ldots & S_{N_t-1}^{(t)} & S_{N_t}^{(t)} & X_1^{(t)} & \ldots & X_{N-N_t}^{(t)} \\ \hline S_2^{(1)} & S_3^{(1)} & \ldots & S_{N_1}^{(1)} & X_1^{(1)} & X_2^{(1)} & \ldots & & X_{N-N_1+1}^{(1)} \\ \vdots & \vdots & & & & & & & \vdots \\ S_2^{(\ell)} & S_3^{(\ell)} & \ldots & \ldots & \ldots & \ldots & \ldots & S_N^{(\ell)} & X_1^{(\ell)} \\ \vdots & \vdots & & & & & & & \vdots \\ S_2^{(t)} & S_3^{(t)} & \ldots & \ldots & S_{N_t}^{(t)} & X_1^{(t)} & X_2^{(t)} & \ldots & X_{N-N_t+1}^{(t)} \\ \vdots & \vdots & & & & & & & \vdots \\ \hline X_{N-N_1}^{(1)} & X_{N-N_1+1}^{(1)} & \ldots & \ldots & \ldots & \ldots & \ldots & \ldots & X_{2N-N_1-1}^{(1)} \\ \vdots & \vdots & & & & & & & \vdots \\ S_N^{(\ell)} & X_1^{(\ell)} & \ldots & \ldots & \ldots & \ldots & \ldots & \ldots & X_{N-1}^{(\ell)} \\ \vdots & \vdots & & & & & & & \vdots \\ X_{N-N_t}^{(t)} & X_{N-N_t+1}^{(t)} & \ldots & \ldots & \ldots & \ldots & \ldots & \ldots & X_{2N-N_t-1}^{(t)} \end{pmatrix} \quad (1)$$

consisting of $N$ stripes. The stripes are created from the $t$ sequences $\mathcal{S}^{(h)}$, $h = 1, \ldots, t$, by shifting the sequences one element to the left in each stripe, and appending a new wild card at the end. With the matrix $\boldsymbol{S}$, our problem is transformed to the problem of finding the minimum $l$ and an appropriate connection polynomial, such that the first $l+1$ columns of $\boldsymbol{S}$ are linearly dependent:

**Problem 3** *Let $\mathcal{S}^{(h)} = \{\tilde{S}_i^{(h)}\}_{i=1}^{N}$, $h = 1, \ldots, t$, be $t$ sequences of length $N = \max\{N_h\}$, where the last $N - N_h$ symbols of every sequence are wild cards, and let $\boldsymbol{S} = (\boldsymbol{s}_1, \ldots, \boldsymbol{s}_N)$ be a matrix created from these sequences pursuant to (1). Then, the* varying length multi-sequence shift-register problem *is transformed into the following problem: Among all possible choices of the wild cards, find the smallest integer $l$, and the polynomial $\sigma(x) = 1 + \sigma_1 x + \cdots + \sigma_l x^l$, $\deg(\sigma(x)) \leq l$, such that the first $l+1$ columns of $\boldsymbol{S} = (\boldsymbol{s}_1, \ldots, \boldsymbol{s}_N)$ are linearly dependent, i.e., such that*

$$\boldsymbol{s}_{l+1} = -\sum_{i=1}^{l} \sigma_i \boldsymbol{s}_{l+1-i} \;. \quad (2)$$

**Remark:** It can be observed, that Problem 3 is just a reformulation of Problem 2'. Hence, the problems 2, 2', and 3 are equivalent.

Problem 3 can be considered as a subclass of the very general linear algebra problem of finding the smallest initial set of linearly dependent columns in a matrix, which is allowed to contain wild cards at the end of each row. Formally, this problem is stated as follows:

**Problem 4** *Let*

$$\boldsymbol{A} = \begin{pmatrix} \boldsymbol{a}_1 & \boldsymbol{a}_2 & \ldots & \boldsymbol{a}_n \end{pmatrix} = \begin{pmatrix} a_{1,1} & a_{1,2} & \ldots & a_{1,N} \\ a_{2,1} & a_{2,2} & \ldots & a_{2,N} \\ \vdots & \vdots & & \vdots \\ a_{M,1} & a_{M,2} & \ldots & a_{M,N} \end{pmatrix}$$

*be an $M \times N$ matrix. The elements $a_{i,j}$ of this matrix are elements from the field $\mathbb{F}$. However, we allow an arbitrary number of successive elements at the end of each row to be wild cards. Then, the problem is formulated as follows: Among all possible choices of the wild cards, find the minimum $l$, and a polynomial $\sigma(x) = 1 + \sigma_1 x + \cdots + \sigma_l x^l$, $\deg(\sigma(x)) \leq l$, such that the first $l+1$ columns in $\boldsymbol{A}$ are linearly dependent, i.e., such that*

$$\boldsymbol{a}_{l+1} = -\sum_{j=1}^{l} \sigma_j \boldsymbol{a}_{l+1-j} \qquad (3)$$

*is satisfied.*

If $\operatorname{rank}(\boldsymbol{A}) = N$ for every choice of the wild cards, then all columns of $\boldsymbol{A}$ are linearly independent and we formally define the solution of the problem to be $l = N$. The polynomial $\sigma(x)$ is meaningless in this case.

The wild cards in $\boldsymbol{A}$ have the same meaning as the wild cards in the sequences, i.e., they are placeholders which can be replaced by arbitrary elements from $\mathbb{F}$. However, unlike in Problem 3, where the wild cards in the rows of $\boldsymbol{S}$ are related to each other, there is no relation between the wild cards in the rows of $\boldsymbol{A}$, so that they can be assigned independently. This shall be illustrated by the following example:

**Example 3** *Consider the binary matrix*

$$\boldsymbol{A} = \begin{pmatrix} 1 & 0 & 1 & 0 & 1 \\ 1 & 0 & X_1 & X_2 & X_3 \\ 1 & 0 & 1 & X_4 & X_5 \\ 0 & 1 & 1 & 1 & 0 \\ 0 & 1 & 1 & 1 & X_6 \\ 0 & 1 & X_7 & X_8 & X_9 \end{pmatrix}$$

*with nine wild cards which can be chosen to be arbitrary elements from $\mathbb{F}_2$. Choosing $X_1 = 1$ and $X_7 = 1$ results in the first three columns of $\boldsymbol{A}$ to be linearly dependent. Hence, a valid solution for Problem 4 is $l = 2$ and $\sigma(x) = 1 + x + x^2$.*

To describe the relation between Problem 2 and Problem 4, we state the following lemma:

**Lemma 1** *A solution for Problem 4 with $\boldsymbol{A} = \boldsymbol{S}$ gives a solution for Problem 2.*

Lemma 1 immediately follows from the structure of the matrix $\boldsymbol{S}$ described by (1), and the equivalence of Problem 2, Problem 2', and Problem 3. Hence, we omit a formal proof.

The fact that we allow wild cards in the matrix $\boldsymbol{A}$ can be interpreted as solving the problem with a set of matrices, and then taking the matrix with the shortest length solution. Particularly in the varying length case, the rightmost elements of some rows become wild cards to accommodate shorter sequences to the matrix structure. This may result in an irregular, i.e., non-triangular arrangement of the wild cards in the matrix $\boldsymbol{S}$.

In the following, we explain the FIA proposed in [2]. Then we demonstrate, that this algorithm may fail to solve Problem 4 if we have such an irregular wild card structure.

From (3) we see, that we can solve Problem 4 by finding the minimum $l$, and a polynomial $\sigma(x)$ such that

$$a_{i,l+1} + \sum_{j=1}^{l} \sigma_j a_{i,l+1-j} = 0, \ \forall \ i = 1, \ldots, M \ .$$

To simplify the notation, we define

$$a^{(i)}(x) = 1 + a_{i,1}x + a_{i,2}x^2 + \cdots + a_{i,N}x^N$$

to be a polynomial composed of the elements of the $i$th row of $\boldsymbol{A}$. Furthermore, we denote by

$$\left[\sigma(x) \cdot a^{(i)}(x)\right]_m = a_{i,m} + \sum_{j=1}^{\deg(\sigma(x))} \sigma_j a_{i,m-j} \qquad (4)$$

the $m$th coefficient of the polynomial $b(x) = \sigma(x) \cdot a^{(i)}(x)$ corresponding to the monomial $b_m x^m$. We call $\left[\sigma(x) \cdot a^{(i)}(x)\right]_m$ the *discrepancy* for the element $a_{i,m}$. With this notation, Problem 4 becomes finding the minimum $l$ and the polynomial $\sigma(x)$ such that

$$\left[\sigma(x) \cdot a^{(i)}(x)\right]_{l+1} = 0, \ \forall \ i = 1, \ldots, M \ . \qquad (5)$$

If wild cards are involved in calculating (5) for some $i$, they always can be chosen such that the corresponding equation is satisfied. Hence, we obtain $\left[\sigma(x) \cdot a^{(i)}(x)\right]_m = 0$, whenever $a^{(i)}$ contains wild cards at some elements $a_{i,j}$ for which $j \leq m$, by implicitly assigning the wild cards to appropriate field elements.

Now we are ready to describe the FIA from [2]. Let $\boldsymbol{D} = (d_{i,j})$ be an $M \times N$ matrix called discrepancy matrix, and let $\Sigma = [\sigma^{\langle 1 \rangle}(x), \sigma^{\langle 2 \rangle}(x), \ldots, \sigma^{\langle N \rangle}(x)]$ be some table to store temporary $\sigma$-polynomials $\sigma^{\langle i \rangle}(x)$ for all columns $i = 1, \ldots, N$ of $\boldsymbol{A}$. Further, let the row index $r$, and the column index $s$ be two integers indexing an element in the matrices $\boldsymbol{A}$ and $\boldsymbol{D}$. With this notation, the FIA is described by Algorithm 1 in pseudo code.

After initializing $\boldsymbol{D} = \boldsymbol{0}$, $\Sigma = [\emptyset]$, $\sigma(x) = 1$, $r = 1$, and $s = 1$, the algorithm starts processing $\boldsymbol{A}$ with the leftmost element in the first row. For this element, the algorithm calculates the discrepancy $d = \left[\sigma(x) \cdot a^{(1)}(x)\right]_1$. If $d = 0$, processing is continued in the next row (i.e., $r$ is incremented by one). If $d \neq 0$, the discrepancy is stored at $d_{1,1}$ in the matrix $\boldsymbol{D}$, and the interim $\sigma$-polynomial $\sigma(x)$ is saved as $\sigma^{\langle 1 \rangle}(x)$ in the table $\Sigma$. Then processing is continued in the first row of the next column. Here, again the discrepancy is calculated. If $d = 0$, the algorithm moves down one row by incrementing $r$, and repeats calculating the discrepancy, until it meets some element for which $d \neq 0$. Generally, if $d \neq 0$ in some row $r$, and there exists a non-zero entry $d_{r,u}$ in $\boldsymbol{D}$, and consequently also an non-empty entry $\sigma^{\langle u \rangle}(x)$ in $\Sigma$, the interim $\sigma$-polynomial $\sigma(x)$ is updated to

$$\sigma(x) \leftarrow \sigma(x) - \frac{d}{d_{r,u}} \sigma^{\langle u \rangle}(x) x^{s-u} . \tag{6}$$

Then, processing is continued in the next row of the same column. If there does not exist any non-zero entry in the $r$th row of $\boldsymbol{D}$, we say that the algorithm *gets stuck*. In this case, $d$ is stored as element $d_{r,s}$ in $\boldsymbol{D}$, and $\sigma(x)$ is memorized as $\sigma^{\langle s \rangle}(x)$ in $\Sigma$. After this, processing is continued in the first row of the next column. It should be pointed out, that by this procedure, we only obtain one non-zero entry in any row of $\boldsymbol{D}$, and consequently $d_{r,u}$ in Equation (6) is always unique. In this way, the FIA moves through $\boldsymbol{A}$ on a zigzag trajectory until it reaches the last row of the matrix.

---

**Algorithm 1:** Fundamental Iterative Algorithm

**input**: $M \times N$ matrix $\boldsymbol{A}$

$\boldsymbol{D} = (d_{i,j}) \leftarrow \boldsymbol{0}$, $\Sigma \leftarrow [\emptyset]$, $\sigma(x) \leftarrow 1$, $s \leftarrow 1$, $r \leftarrow 1$
**while** $(r \leq M)$ **and** $(s \leq N)$ **do**
    $d \leftarrow [\sigma(x) \cdot a^{(r)}(x)]_s$    /∗ discrepancy for $a_{r,s}$ ∗/
    **if** $d \neq 0$ **then**
        **if** $\exists u : d_{r,u} \neq 0$ **then**
            $\sigma(x) \leftarrow \sigma(x) - \frac{d}{d_{r,u}} \sigma^{\langle u \rangle}(x) x^{s-u}$
            $r \leftarrow r + 1$    /∗ move down ∗/
        **else**
            $d_{r,s} \leftarrow d$    /∗ store discrepancy in $\boldsymbol{D}$ ∗/
            $\sigma^{\langle s \rangle}(x) \leftarrow \sigma(x)$    /∗ store current $\sigma(x)$ in $\Sigma$ ∗/
            $s \leftarrow s + 1$    /∗ move right ∗/
            $r \leftarrow 1$    /∗ move to the top ∗/
    **else**
        $r \leftarrow r + 1$    /∗ move down ∗/
$l \leftarrow s - 1$
**output**: $l$, $\sigma(x)$

---

To illustrate how the FIA works, we consider an example for a binary matrix $\boldsymbol{A}$ without wild cards:

**Example 4** *Consider the $6 \times 7$ matrix*

$$\boldsymbol{A} = \begin{pmatrix} 0 & 0 & 1 & 1 & 0 & 1 & 0 \\ 1 & 1 & 1 & 1 & 0 & 0 & 0 \\ 0 & 1 & 1 & 0 & 1 & 0 & 1 \\ 1 & 1 & 1 & 0 & 0 & 0 & 1 \\ 1 & 1 & 0 & 1 & 0 & 1 & 1 \\ 1 & 1 & 0 & 0 & 0 & 1 & 0 \end{pmatrix} .$$

*The FIA starts processing this matrix at the upper left corner with the $\sigma$-polynomial $\sigma(x) = 1$. Since the first element $a_{1,1}$ is zero, the discrepancy is also calculated to be zero. Therefore, the algorithm moves on to the element $a_{2,1}$ for which $d = 1$. Since there is not yet any non-zero entry in the second row of*

*$\boldsymbol{D}$, this discrepancy value is stored in the element $d_{2,1}$ of the matrix $\boldsymbol{D}$, and $\sigma(x)$ is stored in table $\Sigma$, like depicted in the graphical scheme below. After this, the FIA jumps to the first element of the second column and calculates the discrepancy for $a_{1,2}$. Since $d$ is also zero for this element, processing is continued at $a_{2,2}$. Here we have a non-zero discrepancy. However, since there already exists a non-zero element in the second row of $\boldsymbol{D}$, the discrepancy can be turned to zero at this position by using $d_{2,1}$ stored in $\boldsymbol{D}$ and $\sigma^{\langle 1 \rangle}(x)$ stored in $\Sigma$ to calculate $\sigma(x) = 1 + x$ according to (6). After this, processing is continued at the element $a_{3,2}$. Here we again have a non-zero discrepancy but no non-zero entry in the corresponding row of $\boldsymbol{D}$. Consequently, $d$ and $\sigma(x)$ are stored in $\boldsymbol{D}$ and $\Sigma$ respectively, and processing is continued at $a_{1,3}$. In the graphical scheme, the positions at which a modification of $\sigma(x)$ are performed are marked by boxes, positions at which the algorithm gets stuck, and $d$ and $\sigma(x)$ are stored, are marked by circles.*

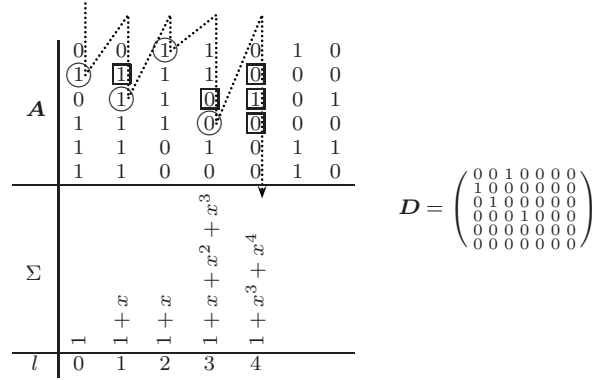

*The FIA continues processing $\boldsymbol{A}$ along the depicted trajectory. Whenever it meets a non-zero discrepancy, it checks whether this discrepancy can be fixed using some previously stored element from $\boldsymbol{D}$ and $\Sigma$. If no previously stored data is available, it jumps to the first element of the next column, until it reaches the element $a_{4,5}$. Here, $\sigma(x)$ is modified by using $d_{4,4}$ and $\sigma^{\langle 4 \rangle}(x)$ from $\boldsymbol{D}$ and $\Sigma$ respectively. After this modification, the discrepancy in the 5th column will be zero for all remaining rows. Hence, the FIA reaches the last row of $\boldsymbol{A}$ in column $s = 5$ and terminates with $l = s - 1 = 4$. Consequently, the first $l = 4$ columns of $\boldsymbol{A}$ are linearly independent. Furthermore, since $\sigma(x) = 1 + x^3 + x^4$ after termination of the FIA, we know, that the sum of the first, the second, and the fifth column will result in the all-zero vector.*

As proved in [2], and demonstrated in the previous example, the FIA works for solving Problem 4 for a matrix $\boldsymbol{A}$ without any wild cards. Now we consider a matrix $\boldsymbol{A}$ with an irregular wild card structure. We have such a structure, for example, if we apply the FIA to the matrix $\boldsymbol{S}$ pursuant to (1), created from multiple sequences of varying length. It is stated in [2], that the FIA can also be applied for this irregular case. However, we show in the following by means of a counter-example, that in general, the FIA does not solve Problem 4, if $\boldsymbol{A}$ has an irregular, i.e., non-triangular wild card structure.

**Example 5** *We again consider the $6 \times 7$ matrix $\boldsymbol{A}$ from Example 4, but now with some independently assignable wild cards at the rightmost positions of the matrix rows:*

$$\boldsymbol{A} = \begin{pmatrix} 0 & 0 & 1 & 1 & 0 & 1 & 0 \\ 1 & 1 & 1 & 1 & 0 & X_1 & X_2 \\ 0 & 1 & 1 & 0 & 1 & 0 & X_3 \\ 1 & 1 & 1 & 0 & X_4 & X_5 & X_6 \\ 1 & 1 & 0 & 1 & 0 & X_7 & X_8 \\ 1 & 1 & 0 & X_9 & X_{10} & X_{11} & X_{12} \end{pmatrix}.$$

*Since the non wild card elements of $\boldsymbol{A}$ coincide with the matrix elements from Example 4, there exists a way to select the wild cards in such a way, that we obtain the same solution as in Example 4 or possibly even another solution with a smaller $l$. Hence, if we apply the FIA to $\boldsymbol{A}$, we expect to get a solution with $l \leq 4$. The graphical scheme below shows the application of Algorithm 1 to the matrix $\boldsymbol{A}$ with wild cards. Whenever a wild card is involved in in calculating $d = [\sigma(x) \cdot a^{(r)}(x)]_s$, the discrepancy is assumed to be zero. This means, that the algorithm implicitly assigns values to the wild cards, such that it yields a zero discrepancy.*

$$\boldsymbol{D} = \begin{pmatrix} 0 & 0 & 1 & 0 & 0 & 0 & 0 \\ 1 & 0 & 0 & 0 & 0 & 0 & 0 \\ 0 & 1 & 0 & 0 & 0 & 0 & 0 \\ 0 & 0 & 0 & 1 & 0 & 0 & 0 \\ 0 & 0 & 0 & 0 & 1 & 0 & 0 \\ 0 & 0 & 0 & 0 & 0 & 0 & 0 \end{pmatrix}$$

*It can be seen, that the FIA behaves exactly in the same as way, as for the case without any wild cards, until it reaches the element $a_{4,5}$. Since we have the wild card $X_4$ at this position, the FIA assumes the discrepancy at this position to be zero. This means, that it implicitly assigns $X_4 = 1$ to the wild card in order to obtain $[\sigma(x) \cdot a^{(4)}(x)]_5 = 0$. Then, the FIA goes ahead to the element $a_{5,5}$ without changing $\sigma(x)$. Here it observes a non-zero discrepancy but no non-zero element in the 5th row of $\boldsymbol{D}$. Consequently, $s$ is incremented to $s = 6$ and processing is continued in the next column until $a_{3,6}$ is reached. At this element, $\sigma(x)$ is updated to $\sigma(x) = 1 + x + x^2 + x^4 + x^5$. Then the FIA reaches the last row of $\boldsymbol{A}$ without further modifying $\sigma(x)$, and terminates with $l = s - 1 = 5$. However, we know from Example 4, that the wild cards in $\boldsymbol{A}$ can be chosen such, that we are able to obtain $l = 4$, and $\sigma(x) = 1 + x^3 + x^4$. Hence, the solution obtained by the FIA is not a valid solution for Problem 4. This shows, that the FIA described in [2] may fail in finding the smallest initial set of linearly dependent columns for matrices with wild cards.*

To understand, why the FIA fails in Example 5, we consider its behavior when it meets a wild card at some position. In such a case, it forces the discrepancy to be zero by implicitly assigning a suitable value to the wild card. If the FIA processes a non wild card element after this, it has no possibility to revise its assignments of the values of the previously processed wild cards. Hence, it continues solving the problem with a smaller set of matrices, which might not contain the smallest length solution. In this sense, the FIA can be considered as a *Greedy Algorithm*. The matrix $\boldsymbol{A}$ used in Example 4 is just an instance of the matrix from Example 5, i.e., we can select the wild cards such that we get $\boldsymbol{A}$ from Example 4. However, after processing $a_{4,5}$, the wild card is implicitly fixed to 1 such that the matrix from Example 4 will not longer be considered as a possible solution.

To fix the problem discovered above, two strategies are conceivable. One possibility is to find an algorithm, which is able to go back and revise its assignments of the wild cards to allow for checking all possible values for each wild card position. The other possibility is to completely avoid the processing of wild cards unless we meet a wild card element $a_{r,s}$ for which all following elements $a_{r+1,s}, \ldots, a_{M,s}$ in the column are also wild cards. In this case we know, that the algorithm will terminate in the same column. Hence, we do not exclude any potential solutions until we find the column at which we terminate and consequently obtain a solution with minimum $l$, regardless of the selection of the wild cards.

Since the first solution seems to be algorithmically quite complex, we propose a modification of the FIA, which is based on the second idea. It turns out, that this solution is quite simple and merely consists of a preprocessing step on the matrix $\boldsymbol{A}$. In the following, we describe this modification and then we formally prove, that the modified algorithm will always succeed in finding the correct solution.

Let $\boldsymbol{A}$ be a $M \times N$ matrix with elements from the field $\mathbb{F}$ which is allowed to contain an arbitrary number of successive wild cards at the rightmost positions of its rows. Now, define by $\text{sort}(\boldsymbol{A})$ an operation on the matrix $\boldsymbol{A}$, such that its rows are sorted from top to bottom to have a non-decreasing number of wild cards. Since permutations of rows do not influence the linear dependency of the columns, solving Problem 4 for the matrix $\check{\boldsymbol{A}} = \text{sort}(\boldsymbol{A})$ yields exactly the same result as solving the problem for the matrix $\boldsymbol{A}$. Just with this simple sorting operation we are able to state a modification of the FIA which we call *Sorted Fundamental Iterative Algorithm* (SFIA). This algorithm overcomes the problem demonstrated by Example 5. Algorithm 2 gives a description of the SFIA in pseudo code.

---

**Algorithm 2:** Sorted Fundamental Iterative Algorithm

**input**: $M \times N$ matrix $\boldsymbol{A}$

$\check{\boldsymbol{A}} \leftarrow \text{sort}(\boldsymbol{A})$
Apply **Algorithm 1** (FIA) to $\check{\boldsymbol{A}}$
**output**: $l$, $\sigma(x)$

---

To demonstrate how the SFIA actually behaves in contrast to the FIA, we consider the following example:

**Example 6** *Let $A$ be the same $6 \times 7$ matrix $A$ already considered in Example 5. We obtain the matrix*

$$\check{A} = \begin{pmatrix} 0 & 0 & 1 & 1 & 0 & 1 & 0 \\ 0 & 1 & 1 & 0 & 1 & 0 & X_3 \\ 1 & 1 & 1 & 1 & 0 & X_1 & X_2 \\ 1 & 1 & 0 & 1 & 0 & X_7 & X_8 \\ 1 & 1 & 1 & 0 & X_4 & X_5 & X_6 \\ 1 & 1 & 0 & X_9 & X_{10} & X_{11} & X_{12} \end{pmatrix}$$

*by sorting the rows of $A$ from top to bottom in such a way, that we have a non-decreasing number of wild cards. Then, Algorithm 2 operates on $\check{A}$ like depicted in the graphical scheme below:*

$$A \quad \begin{matrix} 0 & 0 & 1 & 1 & 0 & 1 & 0 \\ 0 & 1 & 1 & 0 & 1 & 0 & X_3 \\ 1 & 1 & 1 & 1 & 0 & X_1 & X_2 \\ 1 & 1 & 0 & 1 & 0 & X_7 & X_8 \\ 1 & 1 & 1 & 0 & X_4 & X_5 & X_6 \\ 1 & 1 & 0 & X_9 & X_{10} & X_{11} & X_{12} \end{matrix}$$

$$\Sigma \quad \begin{matrix} & & & x^2+x^3 & & & \\ & & & & x^3+x^4 & & \\ & & & 1+x & 1+x^3+x^4 & & \\ 1 & 1 & 1 & & & & \end{matrix}$$

$$l \quad \begin{matrix} 0 & 1 & 2 & 3 & 4 & & \end{matrix}$$

$$D = \begin{pmatrix} 0 & 0 & 1 & 0 & 0 & 0 & 0 \\ 0 & 1 & 0 & 0 & 0 & 0 & 0 \\ 1 & 0 & 0 & 0 & 0 & 0 & 0 \\ 0 & 0 & 1 & 0 & 0 & 0 & 0 \\ 0 & 0 & 0 & 0 & 0 & 0 & 0 \\ 0 & 0 & 0 & 0 & 0 & 0 & 0 \end{pmatrix}$$

*We observe, that unlike Algorithm 1, the SFIA processes the first wild card $X_4$ not until it processed all non wild card elements in the forth column. In this way it ensures, that the polynomial $\sigma(x)$ does not need to be modified after $X_4$ is processed, since all symbols below $X_4$ are also wild cards. Hence, whenever the SFIA processes its first wild card and therefore implicitly fixes its value, the correct solution is already determined, and the algorithm will terminate without further modifying $\sigma(x)$ or $l$. In this way, the greedy nature of the FIA can never exclude all valid solutions from the set of possible solutions.*

Now, we go ahead with proving, that Algorithm 2 solves Problem 4 for any number of wild cards at the last elements of each row of $A$. To provide a self-contained paper and ease understanding, we present the complete chain of proof even though some parts are equivalent to the proofs presented in [2]. For readers already familiar with [2], we explicitly point out, at which points these proofs will fail for certain wild card configurations, and how our modification works in eliminating these problems.

First we show, that the update rule (6) applied to some element in the $r$th row will result in a $\sigma(x)$, which yields zero discrepancy in the $r$th and all previous rows.

**Lemma 2** *Let $\check{a}_{r-1,s}$ and $\check{a}_{r,s}$ be two elements in the sth column of $\check{A}$. Denote by $\sigma_{r-1}(x)$ the $\sigma$-polynomial after processing $\check{a}_{r-1,s}$ such that $[\sigma_{r-1}(x) \cdot \check{a}^{(r-1)}(x)]_s = 0$. Further assume, that we have a non-zero discrepancy $d = [\sigma_{r-1}(x) \cdot \check{a}^{(r)}(x)]_s$, and that there exist a non-zero entry $d_{r,u}$ in the $r$th row and the $u$th column of the discrepancy matrix $D$, and a corresponding entry $\sigma^{\langle u \rangle}(x)$ in the table $\Sigma$. Then,*

$$\sigma_r(x) = \sigma_{r-1}(x) - \frac{d}{d_{r,u}}\sigma^{\langle u \rangle}(x)x^{s-u}, \quad (7)$$

*is such that*

$$[\sigma_r(x) \cdot \check{a}^{(i)}(x)]_s = 0, \ \forall i = 1, \ldots, r \quad (8)$$

*is satisfied. This means, that $\sigma_r(x)$ yields zero discrepancies for the $r$th and all previous rows in the $s$th column.*

The proof for this lemma applies techniques, which are quite similar to the techniques used in [5] to show the correctness of the Berlekamp–Massey algorithm.

*Proof:* By inserting (7) into (8) we obtain

$$[\sigma_r(x) \cdot \check{a}^{(i)}(x)]_s$$
$$= [\sigma_{r-1}(x) \cdot \check{a}^{(i)}(x)]_s - \frac{d}{d_{r,u}}[\sigma^{\langle u \rangle}(x)\check{a}^{(i)}(x)x^{s-u}]_s$$

From (4) we deduce, that

$$[\sigma_r(x) \cdot \check{a}^{(i)}(x)x^p]_j = [\sigma_r(x) \cdot \check{a}^{(i)}(x)]_{j-p},$$

and consequently we can write

$$[\sigma_r(x) \cdot \check{a}^{(i)}(x)]_s$$
$$= [\sigma_{r-1}(x) \cdot \check{a}^{(i)}(x)]_s - \frac{d}{d_{r,u}}[\sigma^{\langle u \rangle}(x)\check{a}^{(i)}(x)]_u$$
$$= \begin{cases} 0 - \frac{d}{d_{r,u}}0 = 0, & 1 \leq i \leq r-1 \\ d - \frac{d}{d_{r,u}}d_{r,u} = 0, & i = r \end{cases},$$

since

$$[\sigma^{\langle u \rangle}(x)\check{a}^{(i)}(x)]_u = \begin{cases} 0, & 1 \leq i \leq r-1 \\ d_{r,u} & i = r \end{cases}$$

due to Algorithm 2. ∎

From Lemma 2 we conclude, that if the algorithm terminates at column $s \leq N$, then the first $s$ columns of $\check{A}$ are linearly dependent. However, to prove that the SFIA solves Problem 4, we do not only need to show that we find an arbitrary initial set of linearly dependent columns, but the smallest one. To be able to do this, we need two further lemmas.

**Lemma 3** *Let $A$ be a matrix which may contain wild cards at the rightmost positions of its rows. Furthermore, let $\check{A} = \mathrm{sort}(A)$. Assume, that the SFIA is processing some wild card element $a_{r,s} = X_i$ in the $s$th column. Then all elements $a_{r+1,s}, \ldots, a_{M,s}$ are also wild cards.*

The statement of Lemma 3 follows immediately from the triangular wild card structure of the sorted matrix $\check{A}$.

**Lemma 4** *Assume that the SFIA is processing some element of $\check{A}$ in the column $s+1$. Then the first $s$ columns of $\check{A}$ are linearly independent. For the case, when $\check{A}$ contains wild cards, it means that it is not possible to assign values from the field $\mathbb{F}$ to them, such that the first $s$ columns of $\check{A}$ are linearly dependent.*

*Proof:* The fact that the FIA processes some element in the $(s+1)$th column implies, that we have a discrepancy matrix $D$ for which each of the first $s$ columns is non-zero. Each of these columns contains exactly one non-zero element $d_{r_j,j}$ in some row $r_j$, such that $r_1 \neq r_2 \neq \cdots \neq r_s$. Hence, we have a submatrix $D_s$ consisting of the first $s$ columns of $D$, which has $s$ rows with only one non-zero element. Consequently, we can bring $D_s$ to an column Echelon-like form $D'_s$ by permuting its columns. From this form we see, that $\text{rank}(D_s) = \text{rank}(D'_s) = s$. To ease understanding of this step, we consider the matrix $D$ from Example 4 and assume, that we are processing some element in column 5. Then we have

$$D_s = \begin{pmatrix} 0 & 0 & 1 & 0 \\ 1 & 0 & 0 & 0 \\ 0 & 1 & 0 & 0 \\ 0 & 0 & 0 & 1 \\ 0 & 0 & 0 & 0 \\ 0 & 0 & 0 & 0 \end{pmatrix} \longrightarrow D'_s = \begin{pmatrix} 1 & 0 & 0 & 0 \\ 0 & 1 & 0 & 0 \\ 0 & 0 & 1 & 0 \\ 0 & 0 & 0 & 1 \\ 0 & 0 & 0 & 0 \\ 0 & 0 & 0 & 0 \end{pmatrix}$$

with $\text{rank}(D_s) = 4$.

Furthermore, we have $s$ polynomials $\sigma^{\langle j \rangle}(x)$ stored in table $\Sigma$ for which we know, that they yield $[\sigma^{\langle j \rangle}(x) \cdot a^{(i)}(x)]_j = 0$ for all rows $i < r_j$, and $[\sigma^{\langle j \rangle}(x) \cdot a^{(i)}(x)]_j \neq 0$ for the row $i = r_j$. Therefore, we can create a $M \times s$ matrix $\widetilde{D}_s = (\tilde{d}_{i,j})$ with $\tilde{d}_{i,j} = [\sigma^{\langle j \rangle}(x) \cdot a^{(i)}(x)]_j$ with the property, that for the column $j$ the first $r_j$ elements coincide with the first $r_j$ elements of the matrix $D_s$. Furthermore we know, that no wild cards have been included in the calculation of these coinciding elements, since if the algorithm would have met any wild card while processing some column $j \leq s$, it follows from Lemma 3 that it would terminate at this column which contradicts the assumption that we are processing column $s+1$. Hence, we also can bring $\widetilde{D}_s$ to a column Echelon-like form $\widetilde{D}'_s$, and conclude from this, that $\text{rank}(\widetilde{D}_s) = s$ regardless of the choices of the wild cards in $\check{A}$. To illustrate this, we again consider the example from above:

$$\widetilde{D}_s = \begin{pmatrix} 0 & 0 & 1 & 0 \\ 1 & 0 & \tilde{d}_{2,3} & 0 \\ \tilde{d}_{3,1} & 1 & \tilde{d}_{3,3} & 0 \\ \tilde{d}_{4,1} & \tilde{d}_{4,2} & \tilde{d}_{4,3} & 1 \\ \tilde{d}_{5,1} & \tilde{d}_{5,2} & \tilde{d}_{5,3} & \tilde{d}_{5,4} \\ \tilde{d}_{6,1} & \tilde{d}_{6,2} & \tilde{d}_{6,3} & \tilde{d}_{5,4} \end{pmatrix} \longrightarrow \widetilde{D}'_s = \begin{pmatrix} 1 & 0 & 0 & 0 \\ \tilde{d}_{2,3} & 1 & 0 & 0 \\ \tilde{d}_{3,3} & \tilde{d}_{3,1} & 1 & 0 \\ \tilde{d}_{4,3} & \tilde{d}_{4,1} & \tilde{d}_{4,2} & 1 \\ \tilde{d}_{5,3} & \tilde{d}_{5,1} & \tilde{d}_{5,2} & \tilde{d}_{5,4} \\ \tilde{d}_{6,3} & \tilde{d}_{6,1} & \tilde{d}_{6,2} & \tilde{d}_{6,4} \end{pmatrix}.$$

Since all elements of $\widetilde{D}_s$ are calculated by $[\sigma^{\langle j \rangle}(x) \cdot a^{(i)}(x)]_j$, it becomes clear by considering (4), that all columns of the matrix $\widetilde{D}_s$ are just linear combinations of the first $s$ columns of $\check{A}$. Hence, the submatrix $\check{A}_s$ composed of these $s$ columns has also $\text{rank}(\check{A}_s) = s$. Consequently, the first $s$ columns of $\check{A}$ are linearly independent, for all possible choices of the wild cards. ∎

Note, that if the FIA operates on an unsorted matrix $A$, all elements of $\widetilde{D}'_s$ may depend on wild cards. Consequently, the upper right corner of the matrix might only be zero for a special choice of these wild cards. Consequently by showing that the rank of $\widetilde{D}'_s$ obtained from the unsorted matrix is equal to $s$, we do not prove that the first $s$ columns are linearly independent regardless of the choice of the wild cards. And in fact, it is not always true as illustrated by Example 5. As we can see there, is possible to assign values to the wild cards in the fifth row such that the first 5 columns are linearly dependent, even though the matrix $\widetilde{D}_s$ resulting from the FIA has rank 5. Hence, Lemma 4 is *not* proved for an unsorted matrix $A$, and indeed, is does not always hold for this case.

Now, we are able to prove the following theorem:

**Theorem 1** *Let $A$ be some $M \times N$ matrix which may contain an arbitrary number of wild cards at the last positions of each row. Applying Algorithm 2 (SFIA) to a matrix $A$ will always solve Problem 4.*

*Proof:* Algorithm 2 terminates, if it reaches the $M$th row in some column $s$ and $[\sigma(x) \cdot a^{(M)}(x)]_s = 0$ is satisfied. From Lemma 2 we know, that $[\sigma(x) \cdot a^{(r)}(x)]_s = 0$ also holds for $1 \leq r < M$ which means, that the first $s$ columns are linearly dependent, and can be linearly combined to the all-zero vector using the corresponding coefficients of $\sigma(x)$. Furthermore, it follows from Lemma 4, that the first $l = s - 1$ columns of $\check{A}$ are linearly independent. Therefore, there cannot exist a smaller $l$, for which the first $l + 1$ columns of $\check{A}$ are linearly dependent. If the algorithm does not terminate before it reaches the last element in the column $s = N$, it increments $s$ and terminates because of the second condition in the while-loop. In this case it obtains $l = N$, which means, that all $N$ columns of $\check{A}$ are linearly independent. Since sorting the rows of a matrix does not influence the linear dependency of its columns, the statements from above also hold for the matrix $A$. ∎

## V. A Modified Iterative Algorithm utilizing the structure of the matrix $S$

In the previous section we described and proved an algorithm for solving Problem 4. However, our actual goal is to find an algorithm to synthesize a linear feedback shift-register for generating multiple sequences, i.e., for solving Problem 2. As already mentioned, this problem can be considered as a special instance of Problem 4 in which we use a matrix $S$ like shown in (1), composed from the $t$ sequences we like to generate. If all these sequences have the same length, the first $t$ rows do not contain any wild cards. Hence, the matrix $S$ is already structured such, that its rows are sorted from top to bottom with a non-decreasing number of wild cards. Consequently, the FIA will always succeed in finding the correct solution for the shift-register synthesis problem. However, if our $t$ sequences have different lengths, some of the $t$ rows in the first stripe already contain wild cards. This means, that $S$ is not longer sorted with respect to a non-decreasing number of wild cards in the rows. Hence, for varying length sequences, we need to apply the SFIA instead of the FIA to generally obtain a valid solution.

From a theoretical point of view, the multi-sequence shift-register synthesis problem for sequences of varying length is solved at this point. However, for practical reasons it is desirable to have an algorithm, which is as efficient as possible. Therefore we would like to utilize the structure of $\check{S}$ to improve the efficiency of the SFIA. Actually, our goal is to find an algorithm, whose algorithmic structure is similar as possible to the Berlekamp–Massey Algorithm. However, unlike [2], we do not aim directly for this goal but introduce an intermediate step, in which we derive a modification of the SFIA, which utilizes the special structure of $\check{S}$. This modification is of minor practical relevance, but it enables us to understand, *derive*, and formally *prove* our final algorithm.

### A. Utilizing the structure of $\check{S}$ for equal length sequences

To understand how we can manage to reduce the complexity of the SFIA, we have to analyze the structure of the sorted matrix $\check{S} = \text{sort}(S)$. For this purpose, we first consider the case, that all $t$ sequences have the same length $N$. The matrix $\check{S}$ consists of $N$ stripes, each containing $t$ rows with the same number of wild cards. Hence, we can decompose $\check{S}$ into

$$\check{S} = \begin{pmatrix} \check{Z}^{[1]} \\ \check{Z}^{[2]} \\ \vdots \\ \check{Z}^{[N]} \end{pmatrix},$$

whereas for $k = 1, \ldots, N$ any $\check{Z}^{[k]} = \left(\check{z}_{i,j}^{[k]}\right)$ is a $t \times N$ matrix whose rightmost part is a block of $t \times (k-1)$ wild cards. We observe, that for $k > 1$, $1 \leq i \leq t$, and $1 \leq j < N$, the element $\check{z}_{i,j}^{[k]}$ in the matrix $\check{Z}^{[k]}$ is equivalent to the element $\check{z}_{i,j+1}^{[k-1]}$ in the previous matrix $\check{Z}^{[k-1]}$. In this context we want to point out, that "equivalent" does not simply mean that the elements $\check{z}_{i,j}^{[k]}$ and $\check{z}_{i,j+1}^{[k-1]}$ have the same value in *one particular* matrix $S$. It rather means, that in *each* matrix $S$, the elements $\check{z}_{i,j}^{[k]}$ and $\check{z}_{i,j+1}^{[k-1]}$ are identical. To emphasize this, we write $\check{z}_{i,j}^{[k]} \equiv \check{z}_{i,j+1}^{[k-1]}$.

Now, assume that Algorithm 2 operates on the $s$th column of the sorted matrix $\check{S}$ and gets stuck at some element $\check{z}_{h,s}^{[k]}$, $k > 1$, because the discrepancy obtained at this position is non-zero, and there is no non-zero element in the corresponding row $t \cdot (k-1) + i$ of the matrix $D$. From the structure of Algorithm 2, and from Lemma 2 we know, that for the current polynomial $\sigma(x)$ at this position all discrepancies for the previous elements are zero. Consequently, Algorithm 2 continues processing at the first element of the column $s + 1$ and calculates zero discrepancies, until it reaches the element $\check{z}_{h,s+1}^{[k-1]}$ in the matrix $\check{Z}^{[k-1]}$. If there is no non-zero element in the corresponding row $t \cdot \max\{0, k-2\} + i$ of the discrepancy matrix $D$, it encounters the same problem as before for the element $\check{z}_{h,s}^{[k]}$. Hence, if $k > 2$ it jumps again to the first element of the next column and gets stuck in $\check{Z}^{[k-2]}$ unless it finds some non-zero element in the corresponding row of $D$.

From these observations we conclude, that if Algorithm 2 meets a scenario as described above, it just moves up and right on a *staircase* from stripe to stripe without modifying $\sigma(x)$ until it finds a non-zero entry in $D$, or reaches the first stripe of $\check{S}$. However, the SFIA performs this movement on a zigzag trajectory jumping from column to column and calculating discrepancies for many elements for which it is already clear, that they are zero. Consequently, the SFIA can easily be optimized such that it straightly moves up along the staircase instead of jumping up along its zigzag trajectory.

Generally we can say, that the SFIA performs two types of movements while processing the matrix $\check{S}$. If it moves down along a column, it modifies the $\sigma$-polynomial whenever it encounters a non-zero discrepancy, and non-zero entries are available in $D$ and $\Sigma$. We call such a movement *downward movement*. If a non-zero discrepancy is encountered, and no non-zero entries are available in $D$ and $\Sigma$, the algorithm cannot continue its downward movement and gets stuck. If this happens, the algorithm moves up along a staircase of identical matrix elements. We call such a movement *staircase movement*, regardless of whether it is performed on a zigzag trajectory or straight along a staircase. During a staircase movement, the $\sigma$-polynomial is not modified. However a staircase movement implicitly causes a prolongation of the shift-register, as soon as the algorithm meets a non-zero entry in $D$ and switches over again to its next downward movement. If the algorithm reaches the top of a staircase without finding an appropriate non-zero entry in $D$, it continues its downward movement in the next column at the first row of the matrix.

To illustrate the observations discussed above, we consider Fig. 3. Here we have a matrix $\check{S}$ consisting of two sequences of length $8$. In this matrix, an exemplary trajectory is sketched, which demonstrates the different types of movements described above. The matrix elements at which the SFIA gets stuck are surrounded by circles, some elements forming a staircase are shaded grey. Downward movements are depicted by solid lines. A staircase movement along a zigzag trajectory (like performed by the SFIA) is depicted by a dotted line whereas a straight movement along the staircase is drawn as dashed line.

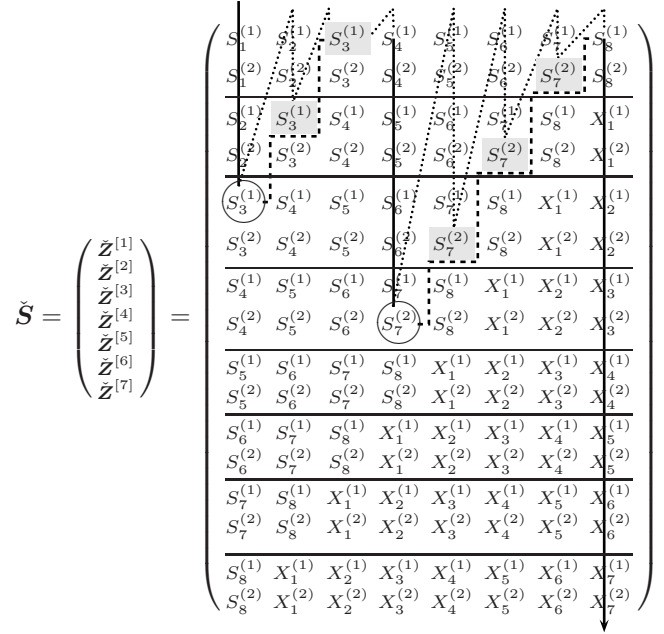

Fig. 3. Movements of the Sorted Fundamental Iterative Algorithm

### B. The varying length case

Now consider the case in which we are actually interested in, i.e., the case of sequences of varying length. Here we have $t$ sequences $\mathcal{S}^{(h)}$, $N_h \leq N$, $h = 1, \ldots, t$ from which at least one has length $N$. Then, as before, the sorted matrix $\check{S}$ consists of $N$ stripes, each containing the same number of wild cards.

Hence, we are again able to decompose $\check{\boldsymbol{S}}$ into

$$\check{\boldsymbol{S}} = \begin{pmatrix} \check{\boldsymbol{Z}}^{[1]} \\ \check{\boldsymbol{Z}}^{[2]} \\ \vdots \\ \check{\boldsymbol{Z}}^{[N]} \end{pmatrix}. \qquad (9)$$

However, unlike in the equal length case, the stripes on the top of $\check{\boldsymbol{S}}$ may now consist of a smaller number of rows, i.e., the matrices $\check{\boldsymbol{Z}}^{[k]}$ are $t_k \times N$ matrices with $t_k \leq t$. Nevertheless, we still have a staircase relation between the matrices $\check{\boldsymbol{Z}}^{[k]}$, but with a possibly decreasing step size at the top of $\check{\boldsymbol{S}}$. This does not change anything at the principle explained above, even though it makes the notation more complicated. To keep the notation as simple as possible, we assume w.l.o.g. that the $t$ sequences are sorted with respect to their lengths, i.e., $N = N_1 \geq N_2 \geq \cdots \geq N_t$. Furthermore, we assume that whenever $\operatorname{sort}(\cdot)$ is applied to $\boldsymbol{S}$, the matrix is sorted in a way, that all rows within a submatrix $\check{\boldsymbol{Z}}^{[k]}$ are ordered such, that a line corresponding to the sequence $\mathcal{S}^{(h)}$ occurs above a line corresponding to $\mathcal{S}^{(g)}$ whenever $h < g$. In this case, an element $\check{z}^{[k]}_{i,j}$ in the matrix $\check{\boldsymbol{Z}}^{[k]}$ is equivalent to the element $\check{z}^{[k-1]}_{i,j+1}$ in the previous matrix $\check{\boldsymbol{Z}}^{[k-1]}$, i.e., $\check{z}^{[k]}_{i,j} \equiv \check{z}^{[k-1]}_{i,j+1}$, whenever $k > N - N_i + 1$, $i < t_k$, and $1 \leq j < N$. The structure of $\check{\boldsymbol{S}}$ is illustrated in Fig. 4 for the case of two sequences, one having length 7, and the other having length 5. Like in Fig. 3, $\check{\boldsymbol{S}}$ consists of seven stripes. However, unlike in the equal length case, the first two stripes only consist of one row. Consequently, we have a staircase structure with stairs of different heights, and some of the staircases do not reach up to the top of the matrix. To illustrate this, the staircases for the elements $S_7^{(1)}$ (solid line) and $S_3^{(2)}$ (dashed line) are sketched. We see, that the two uppermost steps in the staircase of $S_7^{(1)}$ are smaller than the lower steps. Furthermore we observe, that the staircase for $S_3^{(2)}$ does not reach the first stripe in $\check{\boldsymbol{S}}$ but ends at $\check{\boldsymbol{Z}}^{[3]}$.

$$\check{\boldsymbol{S}} = \begin{pmatrix} \check{\boldsymbol{Z}}^{[1]} \\ \check{\boldsymbol{Z}}^{[2]} \\ \check{\boldsymbol{Z}}^{[3]} \\ \check{\boldsymbol{Z}}^{[4]} \\ \check{\boldsymbol{Z}}^{[5]} \\ \check{\boldsymbol{Z}}^{[6]} \\ \check{\boldsymbol{Z}}^{[7]} \end{pmatrix} = \begin{pmatrix} S_1^{(1)} & S_2^{(1)} & S_3^{(1)} & S_4^{(1)} & S_5^{(1)} & S_6^{(1)} & S_7^{(1)} \\ S_2^{(1)} & S_3^{(1)} & S_4^{(1)} & S_5^{(1)} & S_6^{(1)} & S_7^{(1)} & X_1^{(1)} \\ S_3^{(1)} & S_4^{(1)} & S_5^{(1)} & S_6^{(1)} & S_7^{(1)} & X_1^{(1)} & X_2^{(1)} \\ S_1^{(2)} & S_2^{(2)} & S_3^{(2)} & S_4^{(2)} & S_5^{(2)} & X_1^{(2)} & X_2^{(2)} \\ S_4^{(1)} & S_5^{(1)} & S_6^{(1)} & S_7^{(1)} & X_1^{(1)} & X_2^{(1)} & X_3^{(1)} \\ S_2^{(2)} & S_3^{(2)} & S_4^{(2)} & S_5^{(2)} & X_1^{(2)} & X_2^{(2)} & X_3^{(2)} \\ S_5^{(1)} & S_6^{(1)} & S_7^{(1)} & X_1^{(1)} & X_2^{(1)} & X_3^{(1)} & X_4^{(1)} \\ S_3^{(2)} & S_4^{(2)} & S_5^{(2)} & X_1^{(2)} & X_2^{(2)} & X_3^{(2)} & X_4^{(2)} \\ S_6^{(1)} & S_7^{(1)} & X_1^{(1)} & X_2^{(1)} & X_3^{(1)} & X_4^{(1)} & X_5^{(1)} \\ S_4^{(2)} & S_5^{(2)} & X_1^{(2)} & X_2^{(2)} & X_3^{(2)} & X_4^{(2)} & X_5^{(2)} \\ S_7^{(1)} & X_1^{(1)} & X_2^{(1)} & X_3^{(1)} & X_4^{(1)} & X_5^{(1)} & X_6^{(1)} \\ S_5^{(2)} & X_1^{(2)} & X_2^{(2)} & X_3^{(2)} & X_4^{(2)} & X_5^{(2)} & X_6^{(2)} \end{pmatrix}$$

Fig. 4. Staircase structure of $\check{\boldsymbol{S}}$ for sequences of varying length

The fact that we now have stairs of different heights does not prevent us from using the same ideas as before, for the case of equal length sequences. If we are processing some element $z_{h,s}^{[k]}$ corresponding to $\mathcal{S}^{(h)}$ for $k > N - N_h + 1$, and there is no non-zero element in the corresponding row $\sum_{i=1}^{k-1} t_i + h$ of $\boldsymbol{D}$, we can move upwards along the staircase until we find a non-zero element in $\boldsymbol{D}$, or reach the topmost step of a staircase. In the latter case we proceed like before and go ahead with the first element of the following column.

### C. Further improvements due to the staircase structure

The staircase structure of the matrix $\check{\boldsymbol{S}}$ can further be utilized to simplify the SFIA. For this purpose, we focus our considerations again on the staircase movement. Assume that the SFIA gets stuck at some row during a downward movement with a non-zero discrepancy value $d^{(h)}$ corresponding to the sequence $\mathcal{S}^{(h)}$. Then, during the following staircase movement it will inscribe $d^{(h)}$ into the matrix $\boldsymbol{D}$ along an ascending diagonal until it meets a non-zero entry in $\boldsymbol{D}$, or reaches the end of the staircase. Hence, if the SFIA gets stuck the first time in some row $r$ processing an element of the sequence $h$, it will inscribe non-zero entries into all rows of $\boldsymbol{D}$ located above the row $r$, which correspond to the sequence $\mathcal{S}^{(h)}$. If the SFIA gets stuck the next time for the same sequence $\mathcal{S}^{(h)}$ at row $r'$, the following staircase movement will fill all empty rows between row index $r$ and $r'$ corresponding to sequence $\mathcal{S}^{(h)}$. Consequently, the stripe index at which the SFIA may get stuck during a downward movement when processing an element of the sequence $\mathcal{S}^{(h)}$ will increase with any further downward movement. We formalize this observation by the following lemma:

**Lemma 5** *Assume that the SFIA operates on a matrix $\boldsymbol{S}$ and performs a downward movement in column $s$, and that it got stuck at some element $\check{z}_{h,s'}^{[k']}$ in stripe $k'$ during the previous downward movement in column $s'$. Then, the SFIA can get stuck in $\check{z}_{h,s}^{[k]}$ if and only if the discrepancy value $d$ calculated at $\check{z}_{h,s}^{[k]}$ is not zero, and the stripe index is $k > k'$.*

*Proof:* Assume that the SFIA gets stuck in its first downward movement at the element $\check{z}_{h,1}^{[k_1]}$ in the stripe $k_1$. If $k_1 > 1$, the following staircase movement will inscribe non-zero elements to $\boldsymbol{D}$ corresponding to the elements $\check{z}_{h,k_1-i+1}^{[i]}$ into all stripes $\check{\boldsymbol{Z}}^{[i]}$ with $k_1 \geq i \geq 1$, in which the sequence $\mathcal{S}^{(h)}$ is included. Hence, during the next downward movement the algorithm cannot get stuck at some element $\check{z}_{h,\tilde{s}}^{[k_2]}$ as long as $k_2 \leq k_1$, since it always will find a corresponding non-zero entry in $\boldsymbol{D}$ to update the $\Lambda$-polynomial. However, the SFIA will immediately get stuck at a non-zero discrepancy for $\check{z}_{h,\tilde{s}}^{[k_2]}$ if $k_2 > k_1$, because there are no non-zero rows in $\boldsymbol{D}$ corresponding to $\mathcal{S}^{(h)}$ beyond the stripe $k_2$. If the algorithm gets stuck at $\check{z}_{h,\tilde{s}}^{[k_2]}$, $k_2 > k_1$, the following staircase movement will inscribe non-zero entries to the elements of $\boldsymbol{D}$ at all positions corresponding to $\check{z}_{h,k_2-i+\tilde{s}}^{[i]}$, $k_2 \geq i > k_1$. Consequently, there are no all-zero rows corresponding to the sequence $\mathcal{S}^{(h)}$ in the matrix $\boldsymbol{D}$ above the stripe $k_2$, and no non-zero rows beyond this stripe. This induces that the SFIA gets stuck during the following downward movements at the fist stripe beyond $k_2$ at which a non-zero discrepancy is encountered. From this we conclude, that whenever the SFIA gets stuck at some element $\check{z}_{h,s'}^{[k']}$, then after performing a staircase movement there will be no all-zero row for the sequence $\mathcal{S}^{(h)}$ in $D$ above the stripe $k'$ and no non-zero row beyond this stripe. Consequently, the algorithm gets stuck during the next downward movement at element $\check{z}_{h,s}^{[k]}$ if and only if $k > k'$, and a non-zero discrepancy is calculated at this point. ∎

Lemma 5 provides us with a simple and powerful criterion to decide, whether the next processing step has to be a downward or a staircase movement. We do not anymore need to rely on the matrix $D$ to take this decision. Hence, there should be a way to completely dispose the matrix $D$ in our modified algorithm.

### D. An alternative update rule for $\sigma(x)$

To further improve the SFIA, we consider the table $\Sigma$, where we store a temporary $\sigma$-polynomial for each column we process. A polynomial $\sigma^{\langle u \rangle}(x)$ is stored whenever the algorithm processes column $u$ and reaches some row $r = \sum_{i=1}^{k-1} t_i + h$ corresponding to the sequence $\mathcal{S}^{(h)}$ for which it cannot obtain a zero-discrepancy using $d_{r,u}$ and $\sigma^{\langle u \rangle}(x)$. Hence, the table $\Sigma$ may contain several entries corresponding to one and the same sequence. In particular, all entries in $\Sigma$ corresponding to the non-zero entries in $D$, which lie on the same staircase, are temporary $\sigma$-polynomials for the same sequence. If we manage to get by with just one temporary $\sigma$-polynomial for each sequence, we are able to further optimize our algorithm. To show that this is actually possible, we define the polynomial

$$\check{s}^{(i)}(x) = 1 + \check{s}_{i,1} x + \check{s}_{i,2} x^2 + \cdots + \check{s}_{i,N} x^N$$

composed of the elements of the row $i$ of $\check{S}$. Then, we consider the following lemma:

**Lemma 6** *Assume that Algorithm 2 is processing the element $\check{s}_{r,s}$ and obtains $d = [\sigma_{r-1} \cdot \check{s}^{(r)}(x)]_s \neq 0$ using the $\sigma$-polynomial $\sigma_{r-1}(x)$. Further assume, that there exist a row index $r' \geq r$, a column index $s' < s$ and an integer $m$ such that*

$$\begin{aligned} d_{r',s'} &\neq 0 \,, \\ 0 &\leq m < s - s' \,, \\ \check{s}_{r',s'} &\equiv \check{s}_{r,s'+m} \,, \end{aligned} \quad (10)$$

*and that the elements $\check{s}_{r,s}$, $\check{s}_{r',s'}$, and $\check{s}_{r,s'+m}$ correspond to the same sequence $\mathcal{S}^{(h)}$. Then, the polynomial*

$$\sigma_r(x) = \sigma_{r-1}(x) - \frac{d}{d_{r',s'}} \sigma^{\langle s' \rangle}(x) x^{s-s'-m} \quad (11)$$

*satisfies*

$$[\sigma_r(x) \cdot \check{s}^{(i)}(x)]_s = 0, \; \forall i = 1, \ldots, r \,. \quad (12)$$

*This means, that $\sigma_r(x)$ can be used as new interim $\sigma$-polynomial $\sigma(x)$ in Algorithm 2.*

*Proof:* To prove Lemma 6, we insert (11) into the left hand side of (12). In this way, we obtain

$$[\sigma_{r-1}(x) \cdot \check{s}^{(i)}(x)]_s - \frac{d}{d_{r',s'}} [\sigma^{\langle s' \rangle}(x) \cdot \check{s}^{(i)}(x) \cdot x^{s-s'-m}]_s \,. \quad (13)$$

From (4) we deduce, that

$$[\sigma^{\langle s' \rangle} \cdot \check{s}^{(i)}(x) \cdot x^p]_s = [\sigma^{\langle s' \rangle} \cdot \check{s}^{(i)}(x)]_{s-p} \,.$$

Hence, we can rewrite (13) and obtain

$$[\sigma_{r-1}(x) \cdot \check{s}^{(i)}(x)]_s - \frac{d}{d_{r',s'}} [\sigma^{\langle s' \rangle}(x) \cdot \check{s}^{(i)}(x)]_{s'+m} \,. \quad (14)$$

Since we assume, that Algorithm 2 is processing the element $\check{s}_{r,s}$, we know, that

$$[\sigma_{r-1}(x) \cdot \check{s}^{(i)}(x)]_s = \begin{cases} 0 & 1 \leq i < r \\ d & i = r \end{cases} \,. \quad (15)$$

Furthermore, since we assume, that Algorithm 2 already processed the element $\check{s}_{r',s'}$ with $s' < s$ and $r' \geq r$, we have

$$[\sigma^{\langle s' \rangle}(x) \cdot \check{s}^{(j)}(x)]_{s'} = \begin{cases} 0 & 1 \leq j < r' \\ d_{r',s'} & j = r' \end{cases} \,. \quad (16)$$

From (10) we know, that $\check{s}_{r',s'} \equiv \check{s}_{r,s'+m}$. Hence, since Algorithm 2 performs a staircase movement from $\check{s}_{r',s}$ to $\check{s}_{r,s'+m}$, (16) is transformed into

$$[\sigma^{\langle s' \rangle}(x) \cdot \check{s}^{(i)}(x)]_{s'+m} = \begin{cases} 0 & 1 \leq i < r \\ d_{r',s'} & i = r \end{cases} \,. \quad (17)$$

Consequently, by inserting (17), and (15) into (14), the left hand side of (12) becomes

$$\begin{aligned} &[\sigma_{r-1}(x) \cdot \check{s}^{(i)}(x)]_s - \frac{d}{d_{r',s'}} [\sigma^{\langle s' \rangle}(x) \check{s}^{(i+r'-r)}(x)]_{s'} \\ &= \begin{cases} 0 - \frac{d}{d_{r',s'}} 0 = 0, & 1 \leq i \leq r-1 \\ d - \frac{d}{d_{r',s'}} d_{r',s'} = 0, & i = r \end{cases} \,. \end{aligned}$$

This proves Lemma 6. ∎

Now we consider the fact, that $\check{S}$ can be decomposed into stripes according to (9). To accommodate to this stripe structure, we reformulate Lemma 6. For this purpose, we assume, that $\check{s}_{r,s} \equiv \check{z}_{h,s}^{[k]}$ and $\check{s}_{r',s'} \equiv \check{z}_{h,s'}^{[k']}$. Then, $\check{s}_{r,s'+m} \equiv \check{z}_{h,s'+m}^{[k]}$ and $m = k' - k$. Together with the polynomial

$$\check{z}_h^{[k]}(x) = 1 + \check{z}_{h,1}^{[k]} x + \check{z}_{h,2}^{[k]} x^2 + \cdots + \check{z}_{h,N}^{[k]} x^N$$

this gives rise to the following corollary:

**Corollary 7** *Assume, that Algorithm 2 is processing an element $z_{h,s}^{[k]}$ and obtains $d = [\sigma(x) \cdot \check{z}_h^{[k]}(x)]_s \neq 0$ using the interim $\sigma$-polynomial $\sigma(x)$. Further assume, that there exist a stripe index $k' \geq k$, a column index $s' < s$, and a row index $r' = \sum_{i=1}^{k'-1} t_i + h$, such that*

$$\begin{aligned} d_{r',s'} &\neq 0 \,, \\ s' + k' &< s + k \,, \\ \check{z}_{h,s'}^{[k']} &\equiv \check{s}_{r',s'} \,. \end{aligned}$$

*Then, the updating rule*

$$\sigma(x) \leftarrow \sigma(x) - \frac{d}{d_{r',s'}} \sigma^{\langle s' \rangle}(x) x^{s-s'+k-k'} \quad (18)$$

*gives a valid new interim $\sigma$-polynomial.*

Corollary 7 enables us, to modify the SFIA in such a way, that we can keep only one temporary $\sigma$-polynomial $\sigma^{(h)}(x)$ for every sequence $\mathcal{S}^{(h)}$, $h = 1, \ldots, t$, instead of storing the complete table $\Sigma$ with $N$ entries. To do this, we proceed as follows: if we get stuck at the element $\check{z}_{h,s'}^{[k']}$ having $d \neq 0$ with the interim $\sigma$-polynomial $\sigma(x)$, we proceed like the SFIA and assign $\sigma(x)$ to the temporary $\sigma$-polynomial $\sigma^{(h)}(x)$. Together with $\sigma^{(h)}(x)$ we memorize $s'$, $k'$ and $d$ in the auxiliary buffers $s^{(h)}$, $k^{(h)}$, and $d^{(h)}$. To show, that these values are sufficient for further updating, we consider the case, when we next time have $d \neq 0$ for some element $\check{z}_{h,s}^{[k]}$. From Lemma 5 we know, that the SFIA is able to update $\sigma(x)$ only, if $k \leq k'$. Otherwise we get stuck at this point. Hence, if the SFIA is able to update $\sigma(x)$, we always have $k' \geq k$. Furthermore we know, that $s - s' > k' - k$, since otherwise we would be situated above the staircase originating from $\check{z}_{h,s'}^{[k']}$, which would imply $d = 0$. Thus, the conditions of Corollary 7 are satisfied whenever the SFIA is able to update $\sigma(x)$, which in turn means, that we can use the update rule (18) equivalently to the update rule (6) during the downward movements. Consequently, Lemma 5 together with Corollary 7 enables us to completely abandon the matrix $\boldsymbol{D}$.

### E. The Modified Iterative Algorithm (MIA)

Now we propose a *Modified Iterative Algorithm* (MIA), which utilizes the improvements provided by the staircase structure of $\check{s}$, based on the Lemmas 5 and 6. This modified algorithm requires the following auxiliary buffers. For each sequence $\mathcal{S}^{(h)}$, $h = 1, \ldots, t$, we use a buffer $s^{(h)}$ to memorize the column and a buffer $k^{(h)}$ to store the stripe index at which the discrepancy $d^{(h)}$ has been calculated. Then, the MIA is described in pseudo code by Algorithm 3.

Even though the MIA is an improvement compared to the SFIA with respect to various aspects, it still is based on a matrix $\boldsymbol{S}$, and is by far not as elegant as the Berlekamp–Massey algorithm. And indeed, we do not propose the MIA to be used as an improvement of the SFIA. It rather serves us as a theoretical basis to *derive* and *prove* an efficient algorithm with a structure similar to the Berlekamp–Massey algorithm in the next section. Clearly, to be suited for this purpose, we formally need to prove the correctness of Algorithm 3. Therefore, we consider the following theorem:

**Theorem 2** *Let $\mathcal{S}^{(h)}$, $h = 1, \ldots, t$, be $t$ sequences such that $N = N_1 \geq N_2 \geq \cdots \geq N_t$, and let $\boldsymbol{S}$ be a $M \times N$ matrix created from these sequences according to (1). Applying Algorithm 3 (MIA) to the matrix $\boldsymbol{S}$ will always solve Problem 3, and hence also Problem 2.*

*Proof:* Algorithm 3 differs in three aspects from Algorithm 2. The first difference is, that is does not always jump to the top of the column $s + 1$ and continuing processing there, if it gets stuck. Instead, in cases where it hits a staircase, it moves up and right along this staircase, having the same non-zero discrepancy in any step, until it reaches the top of the staircase or a step in which the discrepancy can be fixed. Since all discrepancies above the stairs are known to be zero,

---

**Algorithm 3:** Modified Iterative Algorithm

**input**: $M \times N$ matrix $\boldsymbol{S}$

$\sigma(x) \leftarrow 1$, $s \leftarrow 1$

$\check{\boldsymbol{S}} \leftarrow \text{sort}(\boldsymbol{S})$

$k^{(h)} \leftarrow 0$, $\sigma^{(h)}(x) \leftarrow 0$, $h = 1, \ldots, t$

**for** *each $k$ from 1 to $N$* **do**
    create $\check{\boldsymbol{Z}}^{[k]}$                  /∗ decompose $\check{\boldsymbol{S}}$ into stripes ∗/
    determine $t_k$   /∗ determine number of rows in the stripe ∗/

$k \leftarrow 1$, $h \leftarrow 1$

**while** ($k \leq N$) **and** ($s \leq N$) **do**
    $d \leftarrow [\sigma(x) \cdot \check{z}_h^{[k]}(x)]_s$        /∗ discrepancy for $\check{z}_{r,s}^{[k]}$ ∗/
    **if** $d \neq 0$ **then**
        **if** $k \leq k^{(h)}$ **then**           /∗ move down ∗/
            $\sigma(x) \leftarrow \sigma(x) - \frac{d}{d^{(h)}} \sigma^{(h)}(x) x^{s-s^{(h)}+k-k^{(h)}}$
            $h \leftarrow h + 1$
            **if** $h > t_k$ **then**      /∗ move on to the next stripe ∗/
                $h \leftarrow 1$
                $k \leftarrow k + 1$
        **else**                     /∗ move up along staircase ∗/
            $\tilde{k} \leftarrow k$, $\tilde{s} \leftarrow s$, $\tilde{\sigma}(x) \leftarrow \sigma(x)$
            **if** $k^{(h)} > 0$ **then**
                $\sigma(x) \leftarrow \sigma(x) - \frac{d}{d^{(h)}} \sigma^{(h)}(x) x^{s-s^{(h)}+k-k^{(h)}}$
                $s \leftarrow s + k - k^{(h)}$
                $k \leftarrow k^{(h)}$
            **else**   /∗ staircase ends, jump to the first row ∗/
                $s \leftarrow s + k - N + N_h$
                $k \leftarrow 1$, $h \leftarrow 1$
            $d^{(h)} \leftarrow d$                 /∗ store discrepancy ... ∗/
            $s^{(h)} \leftarrow \tilde{s}$                 /∗ ... column, ... ∗/
            $k^{(h)} \leftarrow \tilde{k}$                 /∗ ... stripe, ... ∗/
            $\sigma^{(h)}(x) \leftarrow \tilde{\sigma}(x)$           /∗ ... and $\sigma(x)$ ∗/
    **else**
        $h \leftarrow h + 1$                    /∗ move down ∗/
        **if** $h > t_k$ **then**      /∗ move on to the next stripe ∗/
            $h \leftarrow 1$
            $k \leftarrow k + 1$

$l \leftarrow s - 1$

**output**: $l$, $\sigma(x)$

---

this movement will result in the same outcome as the zigzag movement of the original SFIA.

The second difference is, that whenever it encounters a non-zero discrepancy $d = [\sigma(x) \cdot \check{z}_h^{[k]}(x)]_s$, it compares the current stripe index $k$ with the stripe index $k^{(h)}$ previously stored for the sequence $h$ and performs a staircase movement whenever $k > k^{(h)}$. It follows from Lemma 5 that this is equivalent to checking the corresponding rows of $\boldsymbol{D}$ for non-zero elements.

The third and most momentous difference to Algorithm 2 is, that we sometimes use different polynomials to update $\sigma(x)$. From Lemma 6 we know, that these alternative polynomials can always be used to update $\sigma(x)$ such that $[\sigma(x) \cdot s'^{(r)}(x)]_s =$ 0 in the row $r$ and all previous rows. Hence, if MIA reaches row $M$ in column $s$ and $[\sigma(x) \cdot s'^{(M)}(x)]_s = 0$ is satisfied after updating $\sigma(x)$ (if necessary), we know that the first $s-1$ columns are linearly independent. Furthermore we observe, that Lemma 4 is applicable to Algorithm 3 without any modification (we only have to replace $A$ by $S$). Consequently we know, that there does not exist a smaller $l = s - 1$, for which the first $l + 1$ columns of $S$ are linearly dependent. ∎

Note, that the MIA and the SFIA sometimes use different temporary polynomials to update $\sigma(x)$. Thus, the two algorithms are not equivalent. And in fact, we do not prove that Algorithm 3 is identical to Algorithm 2 when applied to a matrix $S$. We just show, that it always obtains a valid solution for our problem. So, if this solution is not unique, it is conceivable that the SFIA and the MIA could obtain different results. However, this fact is completely irrelevant for our goal since we are only interested in finding a minimum length solution, we do not to require to obtain a certain one.

## VI. AN EFFICIENT ALGORITHM FOR VARYING LENGTH MULTIPLE SHIFT-REGISTER SYNTHESIS

The SFIA described in Section III is actually merely an algorithm for finding the smallest set of linearly independent columns in a matrix with wild card elements at the rightmost positions of its rows. As shown, it can be used for solving the minimum length shift-register synthesis problem by creating a special structured matrix from the sequences. The improvements of the MIA are achieved by exploiting this special structure. Nevertheless, the MIA is still an algorithm operating on a matrix created from our sequences. However it would be much more natural and also more efficient to find an algorithm which directly operates on the sequences, rather than intricately creating the matrix $S$ before the actual algorithm can be applied. In other words, our goal is to find an algorithm which directly solves Problem 2 without the detour of solving Problem 4 instead. Therefore we propose an efficient shift-register synthesis algorithm whose structure is quite similar to the *Generalized Iterative Algorithm for Multiple Sequences* (GIAMS) described in [2]. However, our algorithm is guaranteed to work also for the case of varying length sequences.

To reach our goal, we consider the MIA and take care that our new algorithm exactly mimics the behavior of the MIA, i.e., all intermediate $\sigma$-polynomials are identical to those created by the MIA. Even more, it is possible to identify a one to one correspondence between the auxiliary variables of the MIA and the new algorithm. In this way, the MIA serves not only as theoretical basis to derive our new algorithm but it enables us also to formally and consistently prove its correctness by showing the equivalence between the processing steps of the MIA and the new algorithm. Regardless of this equivalence, the pseudo code description for our new algorithm given by Algorithm 4 appears even optically much simpler in comparison to Algorithm 3. In fact, the order of complexity of Algorithm 4 is just $t$ times the order of complexity of the Berlekamp–Massey algorithm, which means that calculating the joint linear complexity for $t$ sequences can be performed with the same complexity as calculating the linear complexity for each of this sequences independently.

---

**Algorithm 4:** Efficient Shift-Register Synthesis Algorithm

**input**: $\mathcal{S}^{(h)} = \{S_i^{(h)}\}_{i=1}^{N_h}$, $h = 1, \ldots, t$, $N = \max\{N_h\}$

**begin**

$\quad l \leftarrow 0$, $\sigma(x) \leftarrow 1$
$\quad n^{(h)} \leftarrow N - N_h$, $l^{(h)} \leftarrow 0$, for $h = 1, \ldots, t$
$\quad d^{(h)} \leftarrow 1$, $\sigma^{(h)}(x) \leftarrow 0$, for $h = 1, \ldots, t$
$\quad$ **for** *each $n$ from 1 to $N$* **do**
$\quad\quad$ **for** *each $h$ from 1 to $t$* **do**
$\quad\quad\quad$ **if** $n - l > N - N_h$ **then**   /∗ align to the right ∗/
$\quad\quad\quad\quad d \leftarrow S_{n-N+N_h}^{(h)} + \sum_{j=1}^{l} \sigma_j S_{n-j-N+N_h}^{(h)}$
$\quad\quad\quad$ **if** $d \neq 0$ **then**
$\quad\quad\quad\quad$ **if** $n - l \leq n^{(h)} - l^{(h)}$ **then**
$\quad\quad\quad\quad\quad \sigma(x) \leftarrow \sigma(x) - \frac{d}{d^{(h)}} \sigma^{(h)}(x) x^{n - n^{(h)}}$
$\quad\quad\quad\quad$ **else**
$\quad\quad\quad\quad\quad \tilde{l} \leftarrow l$, $\tilde{\sigma}(x) \leftarrow \sigma(x)$
$\quad\quad\quad\quad\quad \sigma(x) \leftarrow \sigma(x) - \frac{d}{d^{(h)}} \sigma^{(h)}(x) x^{n - n^{(h)}}$
$\quad\quad\quad\quad\quad l \leftarrow n - (n^{(h)} - l^{(h)})$
$\quad\quad\quad\quad\quad l^{(h)} \leftarrow \tilde{l}$, $\sigma^{(h)}(x) \leftarrow \tilde{\sigma}(x)$
$\quad\quad\quad\quad\quad d^{(h)} \leftarrow d$, $n^{(h)} \leftarrow n$

**end**

**output**: $l$, $\sigma(x)$

---

The following theorem formally states the correctness of Algorithm 4:

**Theorem 3** *Let $\mathcal{S}^{(h)}$, $h = 1, \ldots, t$ be $t$ sequences, such that $N = N_1 \geq N_2 \geq \cdots \geq N_t$. Applying Algorithm 4 to this sequences will always solve Problem 2.*

*Proof:* Theorem 3 can be proved by showing that Algorithm 4 is equivalent to Algorithm 3, i.e., that the final result as well as all temporary calculated $\sigma$-polynomials are identical for both algorithms in any step. Therefore we first identify the correspondences between the auxiliary variables in both algorithms. First of all, $l$ is the length of the shift-register in both algorithms, which coincides with the column index $s - 1$ in Algorithm 3, i.e.,

$$s = l + 1. \qquad (19)$$

Furthermore, the variable $n$ in Algorithm 4 corresponds to the index of the rightmost element $S_{n-N+N_h}^{(h)}$ in the subsequence $\{S_i\}_{i=1}^{n-N+N_h}$ currently processed. By examining the structure of the matrix $\check{S}$ we observe, that the shift-register length $l$ and the element index $n$ in Algorithm 4 is related to the stripe index $k$ in Algorithm 3 by

$$k = n - l. \qquad (20)$$

The same holds for the auxiliary variables $l^{(h)}$, $n^{(h)}$ and $k^{(h)}$, $h = 1, \ldots, t$, i.e.,

$$k^{(h)} = n^{(h)} - l^{(h)}. \qquad (21)$$

The polynomial $\sigma(x)$ as well as the auxiliary variables $d^{(h)}$ and $\sigma^{(h)}(x)$ coincide in both algorithms. Now consider the update rule

$$\sigma(x) \leftarrow \sigma(x) - \frac{d}{d^{(h)}}\sigma^{(h)}(x)x^{s-s^{(h)}+k-k^{(h)}}$$

used by Algorithm 3 and the update rule

$$\sigma(x) \leftarrow \sigma(x) - \frac{d}{d^{(h)}}\sigma^{(h)}(x)x^{n-n^{(h)}}$$

used by Algorithm 4. Since we can substitute

$$s - s^{(h)} + k - k^{(h)}$$

by

$$(l+k) - (l^{(h)} + k^{(h)}) = n - n^{(h)} ,$$

the update rules in both algorithms are identical. The discrepancy $d$ is calculated in Algorithm 3 by $d \leftarrow [\sigma(x) \cdot \check{z}_h^{[k]}(x)]_s$. Considering the relation between the stripes $\check{Z}^{[k]}$, and the matrix $\check{S}$ described by (9), we have the following relation:

$$[\sigma(x) \cdot \check{z}_h^{[k]}(x)]_s = z_{h,s}^{[k]} + \sum_{j=1}^{l} \sigma_j z_{h,s-j}^{[k]} =$$

$$= S_{k+s-N+N_h-1} + \sum_{j=1}^{l} \sigma_j S_{k+s-N+N_h-1-j}^{(h)}$$

$$= S_{n-N+N_h} + \sum_{j=1}^{l} \sigma_j S_{n-j-N+N_h}^{(h)} .$$

The latter form is exactly the equation used by Algorithm 4 to calculate the discrepancy $d$.

Now, after we identified the correspondences between the variables in both algorithms, we have to show that they perform matchable operations in any processing step. To do this, we artificially expand the matrix $\check{S}$ as depicted in Fig. 5. In each stripe which has less than $t$ rows, we add dummy rows (shaded gray), such that each of its stripes has an equal number of rows. Furthermore we add a new stripe with the index $k=0$ on top of the matrix, consisting of $t$ dummy lines. Now we assume, that Algorithm 4 performs virtual downward and staircase movements on $\check{S}_{\text{ext}}$. Therefore, we first consider the inner for-loop in the algorithm. Here it is first of all checked whether $n - l > N - N_h$. If this condition is not fulfilled, nothing is done inside the loop. This means, that only the sequence index $h$ and the element index $n$ are updated, but no data processing is performed whenever $n - l > N - N_h$ is not fulfilled. In terms of the matrix notation this condition checks, whether the current stripe index $k$ is larger than the length difference between the longest sequence, and the sequence $\mathcal{S}^{(h)}$. This is the case for all rows of $\check{S}_{\text{ext}}$ but the dummy rows. Hence, Algorithm 4 actually performs data processing only on the non-dummy rows, i.e., the rows of the matrix $\check{S}$. At these rows, it uses identical temporary $\sigma$-polynomials and equivalent values in the auxiliary buffers as we already showed above. Consequently, during the (virtual) downward movements of the Algorithms 3 and 4, both algorithms perform absolutely equivalent data

$$\check{S}_{\text{ext}} = \begin{pmatrix} \check{Z}_{\text{ext}}^{[0]} \\ \check{Z}_{\text{ext}}^{[1]} \\ \check{Z}_{\text{ext}}^{[2]} \\ \check{Z}_{\text{ext}}^{[3]} \\ \check{Z}^{[4]} \\ \check{Z}^{[5]} \\ \check{Z}^{[6]} \\ \check{Z}^{[7]} \end{pmatrix} = \begin{pmatrix} ? & S_1^{(1)} & S_2^{(1)} & S_3^{(1)} & S_4^{(1)} & S_5^{(1)} & S_6^{(1)} \\ ? & ? & ? & S_1^{(2)} & S_2^{(2)} & S_3^{(2)} & S_4^{(2)} \\ S_1^{(1)} & S_2^{(1)} & S_3^{(1)} & S_4^{(1)} & S_5^{(1)} & S_6^{(1)} & S_7^{(1)} \\ ? & ? & S_1^{(2)} & S_2^{(2)} & S_3^{(2)} & S_4^{(2)} & S_5^{(2)} \\ S_2^{(1)} & S_3^{(1)} & S_4^{(1)} & S_5^{(1)} & S_6^{(1)} & S_7^{(1)} & X_1^{(1)} \\ ? & S_1^{(2)} & S_2^{(2)} & S_3^{(2)} & S_4^{(2)} & S_5^{(2)} & X_1^{(2)} \\ S_3^{(1)} & S_4^{(1)} & S_5^{(1)} & S_6^{(1)} & S_7^{(1)} & X_1^{(1)} & X_2^{(1)} \\ S_1^{(2)} & S_2^{(2)} & S_3^{(2)} & S_4^{(2)} & S_5^{(2)} & X_1^{(2)} & X_2^{(2)} \\ S_4^{(1)} & S_5^{(1)} & S_6^{(1)} & S_7^{(1)} & X_1^{(1)} & X_2^{(1)} & X_3^{(1)} \\ S_2^{(2)} & S_3^{(2)} & S_4^{(2)} & S_5^{(2)} & X_1^{(2)} & X_2^{(2)} & X_3^{(2)} \\ S_5^{(1)} & S_6^{(1)} & S_7^{(1)} & X_1^{(1)} & X_2^{(1)} & X_3^{(1)} & X_4^{(1)} \\ S_3^{(2)} & S_4^{(2)} & S_5^{(2)} & X_1^{(2)} & X_2^{(2)} & X_3^{(2)} & X_4^{(2)} \\ S_6^{(1)} & S_7^{(1)} & X_1^{(1)} & X_2^{(1)} & X_3^{(1)} & X_4^{(1)} & X_5^{(1)} \\ S_4^{(2)} & S_5^{(2)} & X_1^{(2)} & X_2^{(2)} & X_3^{(2)} & X_4^{(2)} & X_5^{(2)} \\ S_7^{(1)} & X_1^{(1)} & X_2^{(1)} & X_3^{(1)} & X_4^{(1)} & X_5^{(1)} & X_6^{(1)} \\ S_5^{(2)} & X_1^{(2)} & X_2^{(2)} & X_3^{(2)} & X_4^{(2)} & X_5^{(2)} & X_6^{(2)} \end{pmatrix}$$

Fig. 5. Expansion of the matrix $\check{S}$

processing steps on the same sequence elements modifying the $\sigma$-polynomials in the same way.

Now we consider the staircase movements. Algorithm 3 performs a staircase movement whenever $k > k^{(h)}$. According to (20), and (21), this is equivalent to the condition $n - l > n^{(h)} - l^{(h)}$ in Algorithm 4. First we assume, that the temporary polynomial $\sigma^{(h)}(x)$ is already initialized, i.e., that $\sigma^{(h)}(x) \not\equiv 0$ for some $h$. This implies, that in Algorithm 3 the auxiliary variable $k^{(h)}$ is such that $k^{(h)} > 0$. Consequently, a staircase movement is performed by assigning $s \leftarrow s + k - k^{(h)}$ and $k \leftarrow k^{(h)}$, i.e., by ascending the staircase $k - k^{(h)}$ steps. Due to (19), (20), and (21), this is equivalent to assigning $l \leftarrow n - (n^{(h)} - l^{(h)})$ in Algorithm 4. As shown above, the $\sigma$-polynomial is updated by the same rule in both cases.

Now assume, that we perform a staircase movement for some $h$ for which $\sigma^{(h)}(x) \equiv 0$, and consequently also $k^{(h)} = 0$ in Algorithm 3. On the first glance, the two algorithms behave different in this case. Algorithm 3 will not perform any update of $\sigma(x)$ but moves to the column $s + k - N + N_h$ and performs a downward movement beginning from the first row. Algorithm 4 updates the $\sigma$-polynomial, but since $\sigma^{(h)}(x) = 0$ this update does not effectively change $\sigma(x)$. The staircase movement is performed by updating $l \leftarrow n - N + N_h$. If we again imagine Algorithm 4 virtually operating on the matrix $\check{S}_{\text{ext}}$, this is equivalent to moving right to column $k + s - N + N_h$. However, unlike Algorithm 3, the downward movement is not performed from the first row, but from row $h + 1$ in stripe $k = N - N_h$ if $h < t$, and in row 1 of stripe $N - N_h + 1$ if $h = t$. However, due to the structure of $\check{S}$ and $\check{S}_{\text{ext}}$ respectively, all elements above this point are such that $n - l \leq N - N_h$ holds or they yield a zero discrepancy. Consequently, during its downward movement, Algorithm 3 does not modify $\sigma(x)$ and the auxiliary variables until it reaches the row where Algorithm 4 starts its downward movement. On the other hand, Algorithm 4 skips all rows for which $n - l \leq N - N_h$, i.e., all virtual rows which are not included in $\check{S}$. Hence, both algorithms behave equivalent also in this situation.

From these considerations we conclude, that whenever Algorithm 3 and Algorithm 4 change the $\sigma$-polynomials $\sigma(x)$ and $\sigma^{(h)}(x)$, or the length $l$, they do it in exactly the same way using the elements of the sequences in the same order. In this sense, the two algorithms are equivalent. ∎

To ease understanding of the proof, we illustrate some of the applied arguments by means of the following example:

**Example 7** *Consider the matrices $\check{S}$ and $\check{S}_{ext}$ created from three sequences of length 7, 6 and 5 like depicted in Fig. 6. The matrix $\check{S}$ consists of 7 stripes with the indices $k = 1, \ldots, 7$ and a total of 18 rows. To create $\check{S}_{ext}$, we add a stripe with index $k = 0$ consisting of three dummy rows at the top of this matrix and complement the first stripe by adding two dummy rows and the second stripe by adding one dummy row. In this way, we get a matrix with 8 stripes and a total of 24 rows. The additional rows are marked by hatched boxes.*

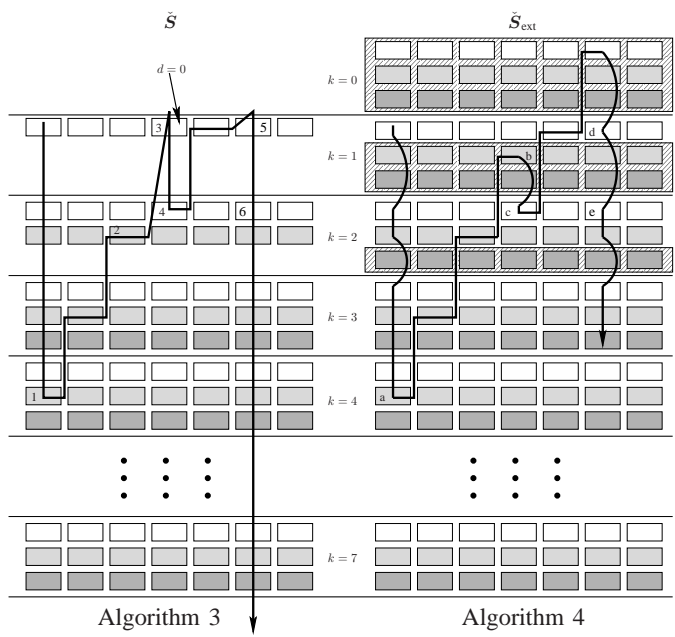

Fig. 6. Equivalence between Algorithm 3 and Algorithm 4

*Now we consider a fictitious trajectory of Algorithm 3 trough the matrix $\check{S}$ and compare it to the virtual trajectory of Algorithm 4, if we imagine it to operate on the matrix $\check{S}_{ext}$. Both algorithms start in the first column at the first row of the stripe $k = 1$ with a downward movement. Assume that Algorithm 3 obtains zero discrepancies, until it gets stuck at the second row of the fourth stripe (marked by "1"). Here it assigns $\sigma^{(2)}(x) \leftarrow 1$. After Algorithm 4 has processed its first element, it immediately hits two rows in a hatched box for which $n - l \leq N - N_h$ holds. Hence it just skips this rows without performing any data processing (depicted by an arc). This means, that Algorithm 4 performs data processing on the same elements as Algorithm 3, until it also gets stuck at the second row of stripe 4 (marked by "a") and assigns $\sigma^{(2)}(x) \leftarrow 1$. At the points "1" and "a" respectively, both algorithms perform a staircase movement.*

*Algorithm 3 stops at the end of the staircase in the second stripe (element marked by "2") and jumps to the first row of the fourth column. Here it starts his next downward movement, at which it encounters a zero discrepancy at the first element (marked by "3"), since the same element has already been processed with a zero discrepancy in the first column of stripe 4. Therefore, Algorithm 3 gets stuck the next time in the second stripe (element marked by "4") and assigns $\sigma^{(1)}(x) \leftarrow 1$, and $d^{(1)} \leftarrow d$. Algorithm 4 ends its staircase movement at stripe $k = N - N_2 = 1$ in the hatched box (element marked by "b") and skips the rows in the hatched box, until it continues its downward movement at stripe 2 (element marked by "c"), where it also gets stuck. It assigns $\sigma^{(1)}(x) \leftarrow 1$ and $d^{(1)} \leftarrow d$ here. Since the element marked by "4' in $\check{S}$, and the element marked by "c" in $\check{S}_{ext}$ coincide, both algorithms again get stuck at the same point. Algorithm 3 processes an extra element (marked by "3"), but since it is clear that this element will yield a zero discrepancy the $\sigma$-polynomial and the auxiliary buffers are identical in both algorithms at the points "4" and "c" respectively.*

*From the point marked by "4" Algorithm 3 performs its next staircase movement which brings it to the top of the matrix where it jumps to the 6th column to start its next downward movement at the point marked by "5". Algorithm 4 performs its staircase movement by jumping to stripe $k = N - N_2 = 0$ which completely consists of dummy rows. Hence, it skips this dummy rows without any data processing and continues its downward movement at the element marked by "d" which coincides with the element marked by "5". Both algorithms update $\sigma(x)$ at this point by using $\sigma^{(1)}(x) \equiv 1$, and $d^{(1)} = d$. At the next elements processed by the algorithms (marked by "6" and "e" respectively), both algorithms again use $\sigma^{(1)}(x) \equiv 1$, and $d^{(1)} = d$ to update $\sigma(x)$ and continue with their downward movement. The element marked by "6" is the last non wild card element in the 6th column such that both algorithms terminate in this column. During the downward movement in column 6, the only elements on the trajectory of Algorithm 4, which differ from the elements on the trajectory of Algorithm 3 are the elements of the dummy rows in the hatched boxes. However, for all of them $n - l \leq N - N_h$ holds, and therefore they are skipped without any data processing such that Algorithm 4 terminates with the same result as Algorithm 3.*

We note, that Algorithm 3 and Algorithm 4 are only equivalent in the sense, that they process the same data and create the same results in all processing steps. However, they are not equivalent with respect to their computational complexity. In a staircase movement during the initialization phase, i.e., when $\sigma^{(h)}(x) \equiv 0$, Algorithm 3 jumps to the top of the matrix and processes several elements which are known to yield a zero discrepancy while Algorithm 4 omits this elements and consequently has a lower computational complexity.

At the end of the day it turns out, that the structure of Algorithm 4 looks quite similar to the structure of the GIAMS from [2]. In fact, our algorithm can be distinguished from the GIAMS by two details. However, these details are non-trivial and crucial extensions to enable the algorithm to work properly with sequences of varying length.

The first difference is, that we initialize the auxiliary buffers $\sigma^{(h)}(x) = 0$ instead of $\sigma^{(h)}(x) = 1$, $h = 1, \ldots, t$. This allows us to prove that Algorithm 4 exactly mimics the processing steps of Algorithm 3 and therefore always solves Problem 2. In contrast to this, the data processing flow in the GIAMS is not always identical to the data processing flow in the FIA. Hence, the correctness of the GIAMS does not immediately follow from the correctness of the FIA in [2]. Moreover, initializing $\sigma^{(h)}(x) = 1$ only works for sequences with equal length, at least in the straightforward way used by [2].

The second difference is the condition $n - l > N - N_h$. In the matrix way of thinking, this condition makes sure, that the rows are always sorted with an increasing number of wild cards. As discussed in Section III, this is essential for the varying length case to obtain the shortest possible solution in any case, even though the assignment of values to the wild cards is done in a "greedy" manner. In the sequence way of thinking, the condition $n-l > N-N_h$ ensures that the shorter sequences are aligned *to the right*. The GIAMS aligns these sequences to the left, which can cause the optimum solution to be excluded from the set of possible solutions.

At this point we also want to note, that the requirements in the Theorems 2 and 3 to have the sequences sorted with respect to their length are introduced only to keep the notation traceable. Algorithm 4 also works, if the input sequences are provided in an arbitrary order. However since a formal proof for this case would require to introduce a quite intricate notation without giving further insight into the topic, this formal proof is omitted here.

## VII. Conclusions

In this paper, we considered the problem of synthesizing the shortest linear feedback shift-register for generating $t$ sequences of varying length over some field $\mathbb{F}$, which is formally stated by Problem 2. Notwithstanding the fact, that it is stated in [2] that Problem 2 can also be solved by the Feng–Tzeng algorithm, we demonstrated that this is generally not true. We identified the specific problem which arises, when the FIA from [2] is applied to sequences of varying length, and explained how this problems can be avoided by an extension of the algorithm (SFIA) which modifies the processing order such that long sequences are always processed before shorter ones. In this way, we always ensure a correct result, notwithstanding of the fact, that we still have the greedy characteristic trait of the original FIA. Furthermore, unlike the original FIA from [2], our versions of the FIA and the SFIA are also applicable for $\text{rank}(\boldsymbol{A}) = N$, i.e., for sequences with maximum linear complexity.

To derive an efficient Berlekamp–Massey like algorithm, we inserted an intermediate step, in which we derived a modification of the SFIA. We proved step by step, that also this modified algorithm (MIA) yields a correct solution to Problem 2, even in cases in which it behaves different from the SFIA. In this way, we obtain a theoretical basis for deriving and proving our final algorithm, since it exactly mimics the data processing flow of the MIA.

The structure of the final algorithm, which we we obtained in this way, is quite similar to the GIAMS from [2]. However, it differs in crucial details from the GIAMS. Our algorithm is different in the order in which the sequences are processed. This is achieved by comparing the difference between the element index $n$ and the shift-register length $l$ to the difference $N - N_h$ and skipping the sequence whenever $n - l \leq N - N_h$. Furthermore, our algorithm differs in the way, in which the auxiliary polynomials $\sigma^{(h)}(x)$ are initialized. The algorithm from [2] uses $\sigma^{(h)}(x) = 1$ to initialize this polynomials while we initialize $\sigma^{(h)}(x) = 0$. This way of initialization ensures the equivalence to the MIA and eliminates initialization problems for the varying length case.

The derivation of our algorithm is performed step by step in a constructive way, starting from the general problem of finding the smallest initial set of linearly dependent columns in a matrix. This step by step approach provides us with a chain of proof, which enables us to formally prove the correctness of Algorithm 4. However, it may also be possible to prove Algorithm 4 a posteriori in a direct way, using the techniques from [3], [18], [19], or others.

## Acknowledgement

The authors would like to thank James L. Massey for valuable discussions and helpful comments on the manuscript.